\documentclass[aps,reprint,superscriptaddress,showkeys,onecolumn
]{revtex4-2}

\usepackage{amsmath}
\usepackage{amssymb}
\usepackage{graphicx,color}
\usepackage{dcolumn}
\usepackage{bm}
\usepackage{hyperref}
\hypersetup{
  colorlinks   = true, 
  urlcolor     = blue, 
  linkcolor    = blue, 
  citecolor   = blue 
}
\allowdisplaybreaks[1]

\def\bs#1{\boldsymbol{#1}}
\def\rmi{{\rm i}}
\def\rmd{{\rm d}}
\def\rme{{\rm e}}
\def\pdiff#1#2{{{\partial #1} \over {\partial #2}}}
\def\diff#1#2{{{\rmd #1} \over {\rmd#2}}}

\def\dsdiff#1#2{{{\rmd^2 #1} \over {\rmd {#2}^2}}}

\def\bs#1{\boldsymbol{#1}}

\begin{document}

\preprint{Preprint \#}

\title[Theoretical modeling of capillary surfers]{Theoretical modeling of capillary surfer interactions on a vibrating fluid bath}

\author{Anand U. Oza}%
\email{oza@njit.edu}
\affiliation{ 
Department of Mathematical Sciences \& Center for Applied Mathematics and Statistics, New Jersey Institute of Technology, Newark, New Jersey 07102, USA}%
 \author{Giuseppe Pucci}
 \affiliation{Consiglio Nazionale delle Ricerche - Istituto di Nanotecnologia (CNR-NANOTEC), Via P. Bucci 33C, 87036 Rende, Italy}
 \author{Ian Ho}%
 \affiliation{School of Engineering, Brown University, 184 Hope Street, Providence, Rhode Island 02912, USA}
\author{Daniel M. Harris}
 \affiliation{School of Engineering, Brown University, 184 Hope Street, Providence, Rhode Island 02912, USA}

\date{\today}

\begin{abstract}
We present and analyze a theoretical model for the dynamics and interactions of ``capillary surfers,” which are millimetric objects that self-propel while floating at the interface of a vibrating fluid bath. In our companion paper~\cite{HoSurfers}, we reported the results of an experimental investigation of the surfer system, which showed that surfer pairs may lock into one of seven bound states, and that larger collectives of surfers self-organize into coherent flocking states. Our theoretical model for the surfers' positional and orientational dynamics approximates a surfer as a pair of vertically oscillating point sources of weakly viscous gravity-capillary waves. 
We derive an analytical solution for the associated interfacial deformation and thus the hydrodynamic force exerted by one surfer on another. Our model recovers the bound states found in experiments and exhibits good quantitative agreement with experimental data. Moreover, a linear stability analysis shows that the bound states are quantized on the capillary wavelength, with stable branches of equilibria separated by unstable ones. Generally, our work shows that self-propelling objects coupled by interfacial flows constitute a promising platform for studying active matter systems in which both inertial and viscous effects are relevant.
\end{abstract}

\keywords{capillary waves, collective motion, active matter}
\maketitle

\section{Introduction}\label{Sec:Intro}

Over the last several decades, there has been significant interest in understanding the physics of so-called ``wet" active matter systems, in which constituents consume energy in order to move through a fluid medium~\cite{Marchetti_Review,Ramaswamy_Review,Gompper_2020}. Such systems are ubiquitous in biology and span the Reynolds-number spectrum. On one end, organisms at the microscale interact through low-Reynolds number (viscous or Stokesian) hydrodynamic interactions~\cite{winkler2018hydrodynamics,Dombrowski2004,Goldstein_Turbulence}. On the other end, schools of fish and flocks of birds generate relatively high-Reynolds number flows in which inertial effects are dominant~\cite{Portugal2014,AshrafPNAS,Wu2011}. Interfacial active systems consist of objects or organisms that self-propel at a liquid-gas interface, and typically exist in an intermediate regime in which both inertial and viscous forces are relevant~\cite{klotsa2019above}. Examples include water-walking insects~\cite{hsieh2004running,bush2006walking,hu2003hydrodynamics,hu2005meniscus}, bio-inspired self-propellers~\cite{yuan2012bio} and self-assembled magnetic swimmers~\cite{snezhko2009self,kokot2017dynamic,sukhov2019}. Prior work has shown that floating solid bodies can self-propel due to the net flow generated by AC electrowetting \cite{yuan2015mechanism}, and that floating water droplets~\cite{pucci2011mutual,ebata2015swimming,pucci2015faraday} and bouncing oil droplets~\cite{Couder2005a,BushOzaROPP} may self-propel across a vibrating fluid bath due to interfacial Faraday waves. Moreover, camphor boats self-propel due to gradients in surface tension~\cite{nagayama2004theoretical,hirose2019} and thus exhibit rich collective behavior~\cite{kohira2001synchronized,suematsu2010collective,heisler2012swarming,ikura2013collective}. 

\begin{figure*}[ht]
  \centering
    \includegraphics[width=1\textwidth]{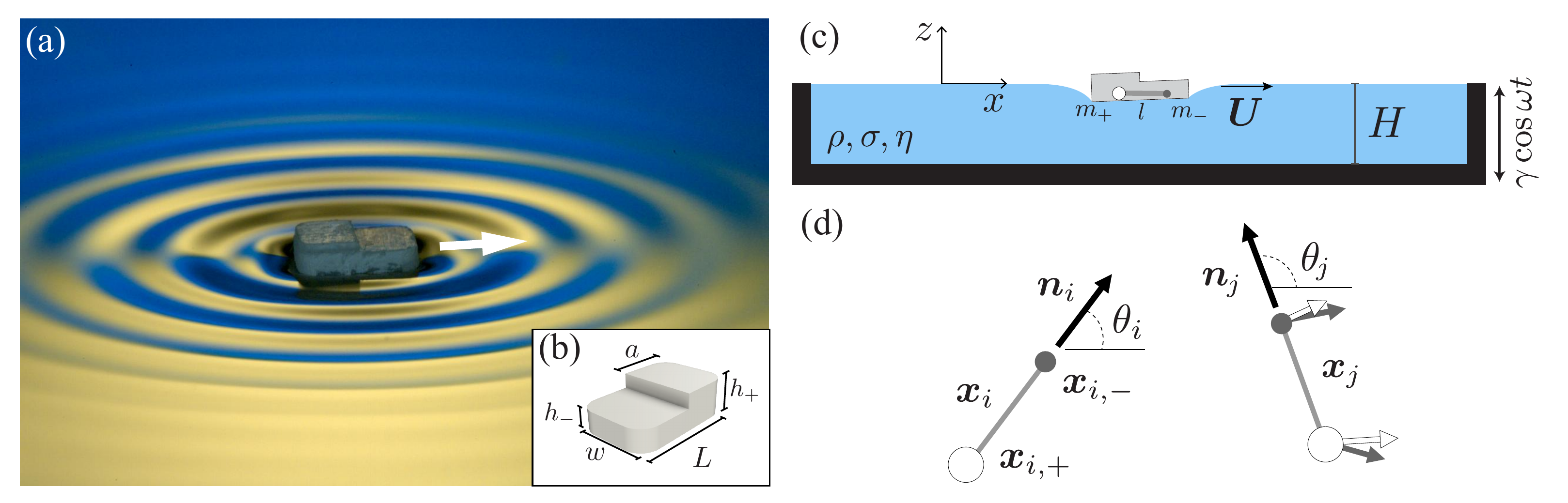}
      \caption{A capillary surfer self-propels on a fluid interface due to its self-generated waves. (a) Oblique wave field visualization, in which colors are obtained from the distorted reflection of a yellow and blue background on the fluid surface. (b) Surfer geometry used in experiments. (c) Side view schematic of the experimental setup (not to scale). The fluid has density $\rho$, surface tension $\sigma$, dynamic viscosity $\eta$ and depth $H$. The theoretical idealization of the surfer is superposed: the surfer is represented as two unequal point masses $m_+$ and $m_-$ connected by a rod of length $l$. (d) Top view schematics of the theoretical model: a surfer with center-of-mass $\bs{x}_i$ experiences a propulsive force $F_p\bs{n}_i$, and its associated point masses located at $\bs{x}_{i,+}$ (white) and $\bs{x}_{i,-}$ (gray) exert capillary wave forces (thick arrows) on the point masses comprising the $j$th surfer.}
        \label{Schematic}
\end{figure*}

In a companion paper~\cite{HoSurfers}, we report the discovery of a new interfacial active system named ``capillary surfers" [Fig.~\ref{Schematic}(a)]. A surfer consists of a millimetric hydrophobic body [Fig.~\ref{Schematic}(b)] that floats on the surface of a vertically vibrating fluid bath of water-glycerol mixture [Fig.~\ref{Schematic}(c)]. All experiments are performed below the Faraday instability threshold, above which subharmonic standing waves spontaneously form at the free surface~\cite{Faraday}. A surfer is front-back asymmetric and thus tilts slightly backwards in equilibrium, with the contact line remaining pinned to the surfer's base perimeter. The vibration of the bath results in the vertical oscillation of the surfer, and the subsequent generation of a radiated, propagating wavefield. The surfer thus moves along its long axis in the direction of its thinner half [Fig.~\ref{Schematic}(a,c)], the velocity being constant in the absence of external perturbations and other surfers. In the following we refer to the front and back of the surfer as the ``bow" and ``stern," respectively. 

For a given surfer geometry, the surfer speed increases with the forcing acceleration and decreases with the forcing frequency [Fig.~S1 in~\cite{HoSurfers}]. Moreover, surfers interact through the wavefields that they generate and thus exhibit novel collective behavior. Specifically, experiments have demonstrated that when pairs of surfers are set into motion towards each other, they may spontaneously arrange into a variety of different bound states [Fig. 2 in~\cite{HoSurfers}]. The system also exhibits multistability: multiple bound states may coexist for the same experimental parameters, and these states are quantized on the capillary length [Fig.~3 in~\cite{HoSurfers}]. Collections of more surfers may self-organize due to their mutual capillary wavefield and exhibit ordered flocking states [Fig.~4 in~\cite{HoSurfers}]. The goal of this paper is to construct and analyze a theoretical model for capillary surfer interactions in order to rationalize the experimental observations.

In order to build such a model, we require a theory for the interfacial deformation induced by capillary-scale floating objects. Approximate expressions for the capillary forces between stationary spherical and cylindrical bodies have been derived~\cite{Chan1980,Kralchevsky1992,Vella2005,Dietrich_1,Danov2005}, and review articles have detailed experimental and theoretical efforts to understand the capillary interactions between bodies trapped at fluid interfaces~\cite{Kralchevsky2000,Oettel2008}. The {\it dynamic} problem, wherein the bodies oscillate at the interface and thus generate a time-dependent wavefield, has received comparatively less attention. Prior work has focused on the deformations generated by relatively large bodies, for which gravitational forces dominate over surface tension~\cite{John1950}. Asymptotic expressions in both the long-~\cite{Ursell1949} and short-wave limits~\cite{Ursell1953,Leppington1972,Leppington1973,RhodesRobinson1982,Simon1985,Keller2013} have been derived. De Corato \& Garbin~\cite{de2018capillary} were the first to derive expressions for small-amplitude capillary waves generated by a periodically oscillating point force at the interface, and the resulting lateral capillary forces experienced by two oscillating point particles. 

Our paper is organized as follows. In \S\ref{Sec:Waves}, we generalize De Corato \& Garbin's work~\cite{de2018capillary} to account for the effects of gravity and weak viscosity. We thus obtain in \S\ref{Sec:Interaction} a formula for the combined static and dynamic forces between two bodies that oscillate at a fluid interface. This formula is used in \S\ref{Sec:SurferModel} to produce a theoretical model for capillary surfers that interact through their collectively generated wave field. In \S\ref{Sec:BoundStates}, we examine the existence and stability of bound states of surfer pairs, and compare our results with experimental data reported in our companion paper~\cite{HoSurfers}. Examples of collective modes exhibited by larger populations of surfers are given in \S\ref{Sec:Collective}. Conclusions and avenues for future work are presented in \S\ref{Sec:Conclusion}.

\section{Weakly viscous linear waves generated by an oscillating point force}\label{Sec:Waves}

In this section, we derive the linear wave field generated by a point force oscillating harmonically on the free surface of a fluid bath. In the experiments~\cite{HoSurfers}, the entire fluid bath is shaken with an acceleration $\gamma\cos(\omega t)$ below the Faraday instability threshold, so we neglect the effects of parametric forcing on the waves. Our analysis generalizes the potential flow model of De Corato \& Garbin~\cite{de2018capillary} by accounting for gravity and weak viscosity. The latter is incorporated by using the approach first given by Lamb~\cite{Lamb1932} and then Dias {\it et al.}~\cite{Dias2008}, wherein viscous corrections to the free surface boundary conditions are derived by assuming that the waves are irrotational and inviscid at leading order, but that dissipation occurs in a viscous boundary layer at the free surface.

\begin{table}
\centering
{\setlength{\extrarowheight}{2pt}
\begin{tabular}{|c|c|c||c|c|c|}
\hline   {\bf Dimensional}  & {\bf Definition} & {\bf Value} & {\bf Dimensionless} & {\bf Definition} & {\bf Value}  \\  
{\bf variable} & {} & {} & {\bf variable} & {} & {} \\ \hline
$\rho$ &  fluid density & 1.175$\times 10^{-3}$ g/mm$^3$ & $\epsilon=2\nu k_c^2/\omega$ & reciprocal Reynolds number & {\it 0.18}  \\ \hline
$\sigma$ &  fluid surface tension & 66 g/s$^2$ & $\beta=1/(k_cl_c)^2$ & wave Bond number & {\it 0.048} \\ \hline
$\eta$ &  fluid dynamic viscosity, & 0.018 g/(mm$\cdot$s), & $k_1$ & wavenumbers in~\eqref{pkDef} & {\it 0.96}$-${\it 0.11} $\rmi$ \\ 
$\nu=\eta/\rho$ & kinematic viscosity & 15.3 mm$^2$/s & $k_2$ & {} & $-${\it 0.47}$-${\it 1.02} $\rmi$ \\ \cline{1-3}
$H$ & fluid depth & 5 mm & $k_3$ & {} & $-${\it 0.48}$-${\it 0.78} $\rmi$ \\ \cline{1-3}
$g$ & gravitational acceleration & 9810 mm/s$^2$ & $k_4$ & {} & $-${\it 31.2}$-${\it 0.36} $\rmi$ \\ \hline
 $f=\omega/2\pi$ & forcing frequency & 20-100 Hz & $\mu_\pm=m_\pm/m$ & mass ratios & 0.6, 0.4 \\ \cline{1-3}
$\gamma=\zeta\omega^2$ & forcing acceleration & 0-3.5 $g$ & $\mu_0=\mu_+-1/2$ & mass offset & 0.1 \\ \hline
$l_c=\sqrt{\sigma/\rho g}$ & capillary length & 2.39 mm & $\text{Bo}=\rho gR^2/\sigma$ & surfer Bond number  & 0.2 \\ \hline
 $L$, & surfer length, & 4.3 mm & $\alpha$ (Eq.~\eqref{FijID_new})& static force coefficient & {\it 0.037} \\ \cline{4-6}
 $l =L/2$, $a$ & half-length, asymmetry & 2.15 mm, 1/2 & $\tilde{l}=lk_c$ & distance between masses & {\it 4.12} \\ \hline
 $k_c=(\rho\omega^2/\sigma)^{1/3}$, & capillary wavenumber, & {\it 1.92} mm$^{-1}$ & $\xi=\gamma/g$ & forcing acceleration & 0-3.5 \\ \cline{4-6}
 $\lambda_c=2\pi/k_c$ & wavelength & {\it 3.28} mm & $\tilde{m}=k_cU\tau_{\text{v}}$ & surfer mass & {\it 2.23} \\ \hline
$w$ & surfer width & 2.7 mm &  $\tilde{F}_{c}=F_{c}/F_{\text{p}}$& dynamic force coefficient & {\it 2.29}$\times${\it 10$^4$}\\ \hline
$\rho_{\text{s}}$ & surfer density & $2.2\times 10^{-3}$ g/mm$^3$ & {} & {} & {} \\ \cline{1-3}
$h_+$, $h_-$ & surfer stern, bow heights & 1.2, 0.8 mm & {} & {} & {} \\ \cline{1-3}
$m$ & surfer mass & 0.026 g & {} & {} & {} \\ \cline{1-3} 
$m_+=aL\rho_{\text{s}}wh_+$  & larger mass & 0.015 g & {} & {} & {} \\ \cline{1-3}
$m_-=(1-a)L\rho_{\text{s}}wh_-$ & smaller mass & 0.01 g & {} & {} & {} \\ \cline{1-3}
$R=L/4$ & surfer effective radius & 1.08 mm & {} & {} & {} \\ \cline{1-3}
$I=m_+m_-l^2/m$ & surfer moment of inertia & 0.028 g$\cdot $mm$^2$ & {} & {} & {} \\ \cline{1-3}
$U$ & surfer free speed & {\it 1.9} mm/s & {} & {} & {} \\ \cline{1-3}
$\tau_{\text{v}}=mH/\eta wL$ & viscous timescale & 0.61 s & {} & {} & {} \\ \cline{1-3}
$F_{\text{p}}=mU/\tau_{\text{v}}$ & propulsive force & {\it 0.08} mm$\cdot$g/s$^2$ & {} & {} & {} \\ \cline{1-3}
$F_{c}=(mg)^2k_c/\sigma$ & dynamic force coefficient & {\it 1.82}$\times${\it 10$^3$} mm$\cdot$g/s$^2$ & {} & {} & {} \\ \hline
\end{tabular}
}
\caption{Variables and parameters appearing in the wave model (\S\ref{Sec:Waves}) and the trajectory equation for surfers (\S\ref{Sec:SurferModel}). Italicized quantities vary with $\gamma$ and $f$, so are reported for the typical combination $f = 100$ Hz and $\gamma = 3.3\,g$.}
\label{tab:param}
\end{table}

Consider an incompressible fluid in an infinite domain $(\bs{x},z)$, where $\bs{x}\in\mathbb{R}^2$ and $z < 0$, $z=0$ being the mean position of the free surface. The fluid has density $\rho$, surface tension $\sigma$ and kinematic viscosity $\nu$, and evolves under the influence of a gravitational acceleration $g$ and an oscillating point force with amplitude $F_0$ and frequency $\omega$. The relevant variables and their characteristic values are listed in Table~\ref{tab:param}. Assuming that the waves are of small amplitude so that the governing equations may be linearized, the free surface height $h(\bs{x},t)$ and velocity potential $\phi(\bs{x},z,t)$ satisfy the system 
\begin{align}
\Delta\phi+\partial_{zz}\phi&=0,\quad z < 0,\quad \bs{x}\in\mathbb{R}^2,\nonumber \\
\partial_t\phi&=-gh+\frac{\sigma}{\rho}\Delta h+2\nu\Delta\phi+\frac{F_0}{\rho}\cos\omega t\,\delta(\bs{x})\quad\text{at}\quad z=0,\nonumber \\
\partial_th&=\partial_z\phi+2\nu\Delta h\quad\text{at}\quad z=0,\nonumber \\
\phi&\rightarrow 0,\quad h\rightarrow 0\quad\text{as}\quad |\bs{x}|,z\rightarrow\infty,\label{ViscGravPotFlow}
\end{align}
where $\Delta = \partial_{xx}+\partial_{yy}$. The first equation enforces the incompressibility of the fluid, while the second and third equations are, respectively, the dynamic and kinematic conditions at the free surface. We solve these equations by writing $\phi(\bs{x},z,t) = \text{Re}\left[\phi_1(\bs{x},z)\rme^{\rmi\omega t}\right]$ and  $h(\bs{x},t) = \text{Re}\left[h_1(\bs{x})\rme^{\rmi\omega t}\right]$, where
\begin{align}
\phi_1(\bs{x},z)=\int_{\mathbb{R}^2}\hat{\phi}_1(\bs{k},z)\rme^{\rmi\bs{k}\cdot\bs{x}}\,\rmd\bs{k}\quad\text{and}\quad h_1(\bs{x})=\int_{\mathbb{R}^2}\hat{h}_1(\bs{k})\rme^{\rmi\bs{k}\cdot\bs{x}}\,\rmd\bs{k}.
\end{align}
The Fourier-transformed quantities $\hat{\phi}_1$ and $\hat{h}_1$ satisfy the algebraic equations
\begin{align}
\partial_{zz}\hat{\phi}_1-|\bs{k}|^2\hat{\phi}_1=0,\quad \rmi\omega\hat{\phi}_1(\bs{k},0)=-g\hat{h}_1-\frac{\sigma}{\rho}|\bs{k}|^2\hat{h}_1-2\nu|\bs{k}|^2\hat{\phi}_1+\frac{F_0}{(2\pi)^2\rho},\quad \rmi\omega\hat{h}_1=\partial_z\hat{\phi}_1(\bs{k},0)-2\nu|\bs{k}|^2\hat{h}_1.\label{FourierEqGrav}
\end{align}
Writing $\hat{\phi}_1(\bs{k},z)=A(\bs{k})\mathrm{e}^{|\bs{k}|z}$, we obtain expressions for $\hat{h}_1$ and $A$:
\begin{align}
\hat{h}_1(\bs{k})=\frac{F_0}{(2\pi)^2}\frac{|\bs{k}|}{\rho(\rmi\omega+2\nu|\bs{k}|^2)^2+\left(\rho g+\sigma|\bs{k}|^2\right)|\bs{k}|}\quad\text{and}\quad A(\bs{k}) = \frac{F_0}{(2\pi)^2}\frac{\rmi\omega+2\nu|\bs{k}|^2}{\rho(\rmi\omega+2\nu|\bs{k}|^2)^2+\left(\rho g+\sigma|\bs{k}|^2\right)|\bs{k}|}.
\end{align}
We are primarily interested in the wave height, so proceed by finding $h_1(\bs{x})$:
\begin{align}
h_1(\bs{x})=\frac{F_0}{2\pi\sigma}\int_0^{\infty}\rmd k\,\frac{k^2}{(\rho/\sigma)(\rmi\omega+2\nu k^2)^2+k/l_c^2+k^3}\mathrm{J}_0(kr)
=\frac{F_0}{2\pi\sigma}\int_0^{\infty}\rmd k\,\frac{k^2}{\epsilon^2k^4+2\rmi \epsilon k^2+k^3+\beta k-1}\mathrm{J}_0(kk_cr),\label{h1IntGrav}
\end{align}
where $|\bs{x}|=r$, and the capillary length $l_c$, capillary wavenumber $k_c$, reciprocal Reynolds number $\epsilon$ and Bond number $\beta$ are defined as, respectively,
\begin{align}
l_c = \sqrt{\frac{\sigma}{\rho g}},\quad k_c = \left(\frac{\rho\omega^2}{\sigma}\right)^{1/3},
\quad \epsilon = \frac{2\nu k_c^2}{\omega},
\quad \text{and}\quad \beta = \frac{1}{(k_cl_c)^2}.\label{ParamDef1}
\end{align}
We note that the weakly viscous wave model~\eqref{ViscGravPotFlow} was derived under the assumption $\epsilon \ll 1$. We also note that $\tanh(k_cH)\approx 1$ where $H$ is the bath depth, which justifies taking the bath to be semi-infinite in the $z$-direction. 

The integrand in Eq.~\eqref{h1IntGrav} can be written as
\begin{align}
\frac{k^2}{P(k)}=\sum_{j=1}^4\frac{A_j}{k-k_j},\quad\text{where } P(k) = \epsilon^2k^4+2\rmi \epsilon k^2+k^3+\beta k-1,\quad A_j = \frac{1}{3+\beta/k_j^2+4\rmi\epsilon /k_j+4\epsilon^2k_j}\label{pkDef}
\end{align}
and $k_j\in\mathbb{C}$ are the roots of $P(k)$. Using the identity~\eqref{BesselID1} in Appendix~\ref{App:dCG}, the integral in Eq.~\eqref{h1IntGrav} may thus be evaluated explicitly:
\begin{align}
h_1(\bs{x}) = \frac{F_0}{2\pi\sigma}\sum_{j=1}^4\frac{1}{3+\beta/k_j^2+4\rmi\epsilon /k_j+4\epsilon^2k_j}\int_0^{\infty}\rmd k\,\frac{\mathrm{J}_0(kk_cr)}{k-k_j}=\frac{F_0}{4\sigma}\sum_{j=1}^4\frac{C_0(-k_jk_cr)}{3+\beta/k_j^2+4\rmi\epsilon/k_j+4\epsilon^2k_j},\label{h1AnalGrav}
\end{align}
where $C_n(r) = \mathrm{H}_n(r)-\mathrm{Y}_n(r)$, $\mathrm{H}_n$ being the $n$th order Struve function and $\mathrm{Y}_n$ the $n$th order Bessel function of the second kind~\cite{abramowitz1948handbook}. The final solution is
\begin{align}
h(\bs{x},t) = \text{Re}\left[h_1(\bs{x})\mathrm{e}^{\rmi\omega t}\right]=\text{Re}[h_1(\bs{x})]\cos\omega t -\text{Im}[h_1(\bs{x})]\sin\omega t.\label{hFinal}
\end{align}
A video of this wavefield is shown in Supplemental Video 1 (left panel).

When implementing the model for interacting capillary surfers (\S\ref{Sec:SurferModel}--\ref{Sec:Collective}), we numerically compute the roots $k_i$ of $P(k)$. However, analytical insight may be obtained by noting that $\epsilon \ll 1$ and $\beta = O(\epsilon^2)$ for a typical value of the forcing frequency $f = 100$ Hz, as shown in Table~\ref{tab:param}. 
One can then show that the roots of $P(k)$ have the following asymptotic expansions in the limit $\epsilon\rightarrow 0$:
\begin{align}
k_1 = 1-\frac{2\rmi\epsilon}{3}+O(\epsilon^2),\quad k_2 = -\varsigma+O(\epsilon),\quad k_3 = -\bar{\varsigma}+O(\epsilon),\quad k_4 = -\frac{1}{\epsilon^2}+O\left(\frac{1}{\epsilon}\right),\label{rootApprox}
\end{align}
where $\varsigma = \rme^{\rmi\pi/3}$. 
Since $\mathrm{H}_0(x)$ and $\mathrm{Y}_0(x)$ both vanish as $x\rightarrow\infty$, we thus obtain the following approximation to Eq.~\eqref{h1AnalGrav}, valid in the regime $0<\epsilon \ll 1$, $r \gg \epsilon^2$:
\begin{align}
h_1(\bs{x}) &\approx \frac{F_0}{12\sigma}\left(\eta_1(r)+2\,\text{Re}\left[\eta_2(r)\right]\right),
\quad\text{where}\quad \eta_1(r)=C_0\left[\left(-1+\frac{2\rmi\epsilon}{3}\right)k_cr\right]\quad\text{and}\quad \eta_2(r)=C_0\left[\varsigma k_cr\right].\label{h1Approx2}
\end{align}
While the real part of $\eta_2(r)$ decays monotonically in $r$, $\eta_1(r)$ decays while oscillating on roughly the capillary wavelength $\lambda_c$. 

\begin{figure}[ht]
\begin{center}
\includegraphics[width=1\textwidth]{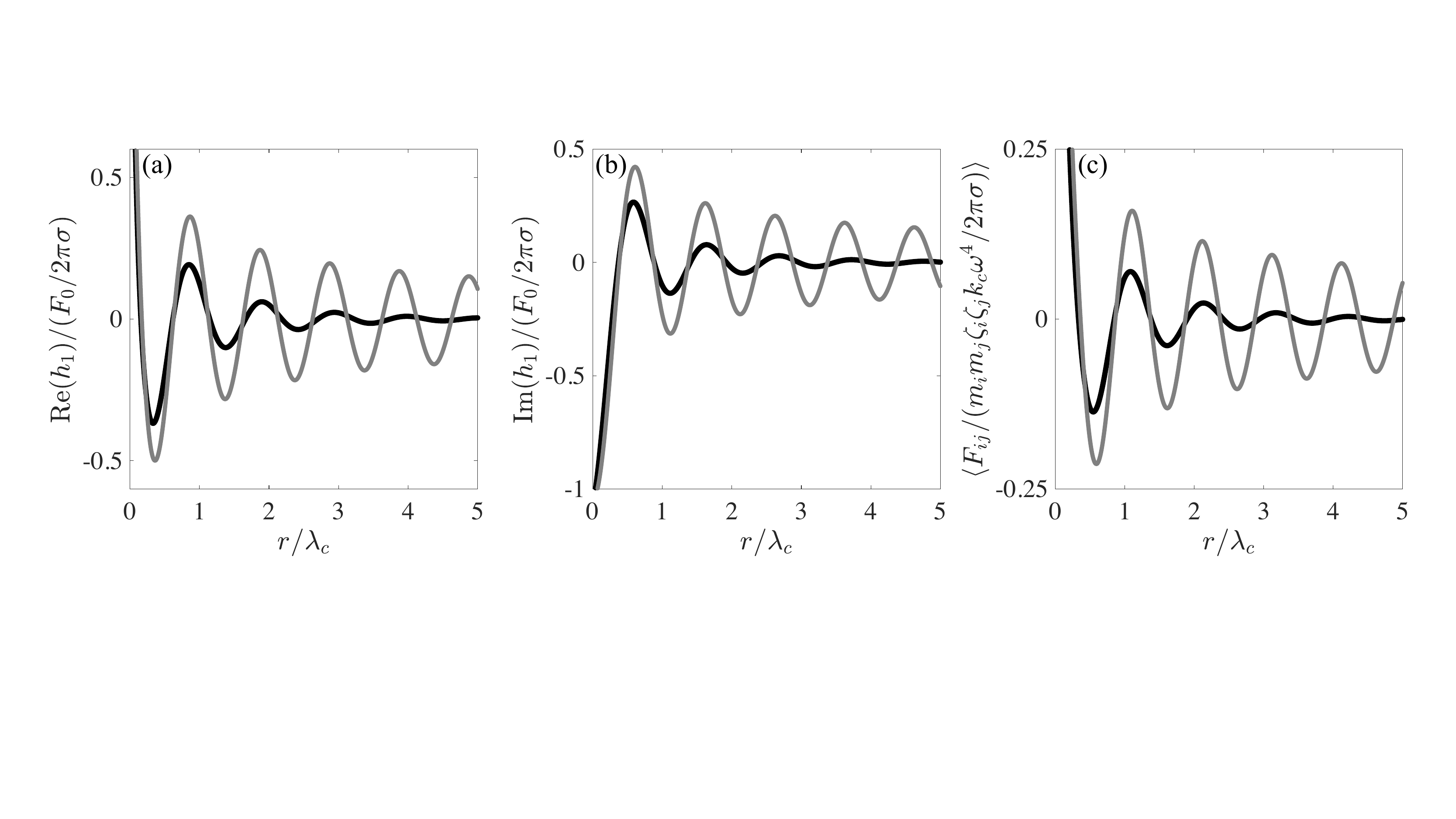}
 \end{center}
 \caption{Real (a) and imaginary (b) parts of the wave height given in Eq.~\eqref{h1AnalGrav} (black curves) are compared against the solution in Eq.~\eqref{h1pm} (gray curves), the latter of which neglects viscous and gravitational effects. Similarly, panel (c) shows the associated force in Eq.~\eqref{ViscGravForce} (black curve), as compared against Eq.~\eqref{dCGForce} (gray curve). The parameters correspond to those given in Table~\ref{tab:param}, with forcing frequency $f = 100$ Hz.
}
 \label{Fig:ViscGrav}
 \end{figure}

In Appendix~\ref{App:dCG}, we derive the solution $h_1(\bs{x})$ in the absence of gravity and viscosity. The derivation closely follows that of De Corato \& Garbin~\cite{de2018capillary}; the significant difference is that we impose the Sommerfeld radiation condition~\eqref{SommerfeldCond}, which enforces that waves propagate outward from the source, while De Corato \& Garbin use a reflecting boundary condition at infinity (see Eq. (2.7) in~\cite{de2018capillary}) and thus obtain a standing waveform. Figure~\ref{Fig:ViscGrav}(a)--(b) show a comparison between the weakly viscous result~\eqref{h1AnalGrav} and the inviscid result $h_1^+(\bs{x})$ in~\eqref{h1pm}.
We observe that, for the typical parameter regime explored in experiments, the inclusion of viscous effects causes $h_1(r)$ to decay faster than its inviscid counterpart. A more detailed discussion of the far-field behavior of $h_1$ is given in Appendix~\ref{App:FarField}. We also note that, since the waves generated by a surfer of mass $m$ have amplitude $A\approx F_0/4\pi\sigma$ where $F_0\approx m\gamma$ (see \S\ref{SSec:Dynamic}), the ratio $A/\lambda_c\approx 0.1-0.3$ over the range $f=20-100$ Hz for the largest value of the forcing acceleration considered, $\gamma=3.5g$, which validates the small-amplitude approximation made in Eq.~\eqref{ViscGravPotFlow}.

\section{Interaction force between a pair of objects oscillating on a fluid interface}\label{Sec:Interaction}

In Section~\ref{Sec:SurferModel}, we propose equations of motion for a collection of interacting surfers. Assuming the interactions to be pairwise, such a framework requires a model for the force between two surfers. To our knowledge, there does not exist an analytical expression for the capillary force between two finite-sized objects oscillating on a fluid interface. For this reason, we make the following simplifying approximations: the ``static" part of the force, induced by the object's weight, is approximated by treating each object as a floating disc. This force is responsible for the so-called ``Cheerios effect"~\cite{Vella2005}, which causes floating objects to clump together. The ``dynamic" part of the force, induced by the object's oscillation on the fluid interface, is approximated by treating each object as an oscillating point particle, which is the problem we solved in \S\ref{Sec:Waves}. This approximation is expected to be valid when the distance between surfers is much larger than the surfer's length $L$.

\subsection{Static force between floating discs}\label{SSec:Static}
We proceed by calculating the interfacial deformation $h_{\text{s}}(r)$ due to a floating disc of radius $R$ and mass $m$ at rest at a depth $\delta$ below the undisturbed free surface. The Young-Laplace equation with Dirichlet boundary conditions,
\begin{align}
\Delta h_{\text{s}}\equiv h_{\text{s}}^{\prime\prime}+\frac{1}{r}h_{\text{s}}^{\prime}=\frac{h_{\text{s}}}{l_c^2},\quad r > R,\quad h_{\text{s}}(R)=-\delta,\quad h_{\text{s}}\rightarrow 0\quad\text{as}\quad r\rightarrow\infty,
\end{align}
has the solution~\cite{Vella2005}
\begin{align}
h_{\text{s}}(r)=-\delta\frac{\mathrm{K}_0(r/l_c)}{\mathrm{K}_0(R/l_c)},\label{hStatic}
\end{align}
where $\mathrm{K}_0$ is the modified Bessel function of the second kind of order zero. The depth $\delta$ is calculated by balancing the disc's weight $mg$ against the buoyancy force $\rho g\pi R^2\delta$ and the vertical component $2\pi R\sigma\sin\theta$ of the surface tension force, $\theta$ being the contact angle of the fluid with the disc: 
\begin{align}
mg=\rho g\pi R^2\delta+2\pi R\sigma\sin\theta.
\end{align}
Assuming $\theta \ll 1$, so that $\sin\theta\approx \tan\theta\approx h^{\prime}(R)$, we obtain
\begin{align}
\delta=\frac{mg}{\pi\sigma}\frac{\mathrm{K}_0(\sqrt{\text{Bo}})}{\text{Bo }\mathrm{K}_0(\sqrt{\text{Bo}})+2\sqrt{\text{Bo}}\,\mathrm{K}_1(\sqrt{\text{Bo}})},\label{deltaCalc}
\end{align}
where we use the fact that $\text{K}_0^{\prime}=-\text{K}_1$. The force $\bs{F}_{ij}^{\text{s}}$ on a disc of mass $m_i$ at $\bs{x}=\bs{x}_i$ due to the static deformation generated by a disc of mass $m_j$ at $\bs{x}=\bs{x}_j$ is
\begin{align}
\bs{F}_{ij}^{\text{s}}\approx -m_ig\bs{\nabla}h_{\text{s}}(r_{ij})=\frac{m_im_jg^2}{\pi\sigma l_c}\frac{\mathrm{K}_1(r_{ij}/l_c)}{\text{Bo }\mathrm{K}_0(\sqrt{\text{Bo}})+2\sqrt{\text{Bo}}\,\mathrm{K}_1(\sqrt{\text{Bo}})}\hat{\bs{x}}_i^j,\quad\text{where}\quad r_{ij} = |\bs{x}_i-\bs{x}_j|\quad\text{and}\quad \hat{\bs{x}}_i^j=\frac{\bs{x}_j-\bs{x}_i}{r_{ij}},\label{FStatic}
\end{align}
assuming that the distance between the discs is much bigger than the capillary length, $r_{ij} \gg l_c$. Note that this force is always attractive, and decays exponentially with the distance between the objects.

\subsection{Dynamic force between oscillating point particles}\label{SSec:Dynamic}

The results of \S\ref{Sec:Waves} can readily be used to compute the force between two point particles with positions $(\bs{x}_i,z_i)$ and $(\bs{x}_j,z_j)$. Assuming that the particles oscillate on the fluid interface with the same phase, $\ddot{z}_j=-\zeta_j\omega^2\cos\omega t$, we now compute the time-averaged force $\bs{F}_{ij}^{\text{d}}$ on particle $i$ due to the deformation generated by particle $j$, their locations being $\bs{x}_i$ and $\bs{x}_j$, respectively. Defining $h_j(\bs{x},t)\equiv h(\bs{x}-\bs{x}_j,t)$, where $h$ is defined in Eq.~\eqref{hFinal}, we obtain
\begin{align}
\boldsymbol{F}_{ij}^{\text{d}} =\langle m_i\ddot{z}_i\bs{\nabla}h_j(\bs{x}_i,t)\rangle=\frac{m_im_j\zeta_i\zeta_j\omega^4}{24\sigma}k_c\sum_{n=1}^4\text{Re}\left[ k_n\frac{\mathrm{H}_{-1}(-k_nk_cr_{ij})+\mathrm{Y}_1(-k_nk_cr_{ij})}{1+\beta/3k_n^2+(4/3)\mathrm{i}\epsilon/k_n+(4/3)\epsilon^2k_n}\right]\hat{\boldsymbol{x}}_i^j,\label{ViscGravForce}
\end{align}
where $\langle\cdot\rangle$ denotes a time average over the oscillation period $2\pi/\omega$, and we use the facts that $\mathrm{Y}_0^{\prime} = -\mathrm{Y}_1$ and $\mathrm{H}_0^{\prime} = \mathrm{H}_{-1}$. A plot of the dynamic force~\eqref{ViscGravForce} is shown in Fig.~\ref{Fig:ViscGrav}(c): while it is attractive when the particles are close together, $r\ll \lambda_c$, it differs from the static force~\eqref{FStatic} in that it oscillates between attractive and repulsive as $r$ increases. Note also that Eq.~\eqref{ViscGravForce}, which incorporates the effects of viscosity, decays much faster than its inviscid counterpart~\eqref{dCGForce}, which was derived by De Corato \& Garbin~\cite{de2018capillary}. 

\section{Trajectory equations for capillary surfers}\label{Sec:SurferModel}

We proceed by constructing the equations of motion for a collection of interacting surfers, the relevant variables being listed in Table~\ref{tab:param}. Consider a surfer with the ``boat" geometry shown in Fig.~\ref{Schematic}(b), with length $L$, width $w$, asymmetry $a$, stern (bow) heights $h_+$ ($h_-$), mass density $\rho_{\text{s}}$, and mass $m$, floating on the free surface of a fluid bath oscillating with acceleration $\gamma$ and frequency $\omega$. Since there does not exist an analytical expression for the force between two surfers oscillating on a fluid interface, we model each surfer as a pair of masses chosen to represent the surfer's asymmetric mass distribution in experiments, $m_+=La\rho_{\text{s}}wh_+$ and $m_-=L(1-a)\rho_{\text{s}}wh_-$ [Fig.~\ref{Schematic}(c)]. These masses are assumed to be non-rotating and connected by a rigid massless rod of length $l=L/2$. For the ``static" part of the force, induced by the surfer's weight, we treat each mass as a disc, and use Eq.~\eqref{FStatic} to approximate the interaction force between two such discs. For the ``dynamic" part of the force, induced by the surfer's oscillation on the fluid interface, we treat each mass as a point particle and use Eq.~\eqref{ViscGravForce} to approximate the interaction force between two point particles. 

We describe the trajectory of the $i$th surfer by its center of mass $\bs{x}_i(t)\in\mathbb{R}^2$ and orientation (unit) vector $\bs{n}_i(t)$, which points from $m_+$ to $m_-$ [Fig.~\ref{Schematic}(d)]. The masses are located at $\bs{x}_{i,\pm}=\bs{x}_i\mp\mu_{\mp}l\bs{n}_i$, where $\mu_{\pm}=m_{\pm}/m$. Each mass moves in response to two forces: wave forces $\bs{F}_{\pm}$, time-averaged over the forcing period $2\pi/\omega$ of the bath, and drag forces $-D_{\pm}\dot{\bs{x}}_{i,\pm}$ due to the viscous shear stress underneath each mass. The equations of motion are thus 
\begin{align}
m_{\pm}\left(\ddot{\bs{x}}_i\mp \mu_{\mp}l\ddot{\bs{n}}_i\right)+D_{\pm}\left(\dot{\bs{x}}_i\mp \mu_{\mp}l\dot{\bs{n}}_i\right)=\bs{F}_{\pm}.
\label{eq:Newton}
\end{align}
We now assume that $D_{\pm}=m_{\pm}/\tau_{\text{v}}$, where $\tau_{\text{v}}=mH/\eta wL$ is the viscous timescale obtained by computing the shear stress due to a locally fully-developed Couette flow on the underside of the surfer. For the sake of simplicity, we neglect the influence of hydrodynamic interactions between the two masses on the values of the drag coefficients. Adding the two equations in Eq.~\eqref{eq:Newton}, we obtain the trajectory equation for the center of mass
\begin{align}
m\left(\ddot{\bs{x}}_i+\frac{1}{\tau_{\text{v}}}\dot{\bs{x}}_i\right)=\bs{F}_++\bs{F}_-.\label{XCM2}
\end{align}
To model the rotational dynamics, we take the cross product of the first equation in Eq.~\eqref{eq:Newton} with $-\mu_-l\bs{n}_i$, the second equation with $\mu_+l\bs{n}_i$, and add the two resulting equations:
\begin{align}
\left[m_-(\mu_+l)^2+m_+(\mu_-l)^2\right]\bs{n}_i\times\ddot{\bs{n}}_i+\frac{1}{\tau_{\text{v}}}\left[m_-(\mu_+l)^2+m_+(\mu_-l)^2\right]\bs{n}_i\times\dot{\bs{n}}_i=l\bs{n}_i\times\left[\mu_+\bs{F}_--\mu_-\bs{F}_+\right].
\label{eq:adds}
\end{align}
Writing $\bs{n}_i=(\cos\theta_i,\sin\theta_i)$, Eq.~\eqref{eq:adds} reduces to
\begin{align}
I\ddot{\theta}_i+\frac{I}{\tau_{\text{v}}}\dot{\theta}_i=l\bs{n}_i\times\left(\mu_+\bs{F}_--\mu_-\bs{F}_+\right),
\label{Orient1}
\end{align}
where $I = m_+(\mu_-l)^2+m_-(\mu_+l)^2 =  \mu_+\mu_-ml^2$ is the moment of inertia in the plane. 

The wave forces $\bs{F}_{\pm}$ may be decomposed into three terms: a propulsive force due to radiation pressure $(F_{\text{p}}/2)\bs{n}_i$, an attractive capillary force~\eqref{FStatic} due to the surfer's weight, and a dynamic wave force~\eqref{ViscGravForce} due to the interfacial waves generated by the surfers. The latter two are expressed as a linear superposition of the forces generated by all other surfers, as shown schematically in Fig.~\ref{Schematic}(d). We thus obtain the trajectory equations
\begin{align}
m\ddot{\bs{x}}_i+\frac{m}{\tau_{\text{v}}}\dot{\bs{x}}_i
&=F_{\text{p}}\bs{n}_i+F_{c}\sum_{p,q=\pm 1}\mu_{p}\mu_{q}\sum_{j\neq i}\Phi\left(k_c|\bs{x}_{j,q}-\bs{x}_{i,p}|\right)\frac{\bs{x}_{j,q}-\bs{x}_{i,p}}{|\bs{x}_{j,q}-\bs{x}_{i,p}|},\nonumber \\
I\ddot{\theta}_i+\frac{I}{\tau_{\text{v}}}\dot{\theta}_i
&=-lF_{c}\sum_{p,q=\pm 1}\mu_{p}\mu_{q}\sum_{j\neq i}p\mu_{-p}\Phi\left(k_c|\bs{x}_{j,q}-\bs{x}_{i,p}|\right)\bs{n}_i\times\frac{\bs{x}_{j,q}-\bs{x}_{i,p}}{|\bs{x}_{j,q}-\bs{x}_{i,p}|},
\label{eq:dimmodel}
\end{align}
where 
$F_{c}=(mg)^2k_c/\sigma$ is the capillary force coefficient. The interaction force $\Phi$ is obtained by adding Eqs.~\eqref{FStatic} and~\eqref{ViscGravForce}, where we assume that the surfer oscillation amplitudes $\zeta_i$ are equal to the forcing amplitude $\gamma/\omega^2$ of the bath:
\begin{align}
\Phi(r) &= \alpha f_{\text{s}}(r)+\frac{\xi^2}{24}f_{\text{d}}(r),\quad \text{where}\quad\xi = \frac{\gamma}{g},\quad \alpha=\frac{\sqrt{\beta}}{\pi\text{ Bo }\left(\mathrm{K}_0(\sqrt{\text{Bo}})+2\mathrm{K}_1(\sqrt{\text{Bo}})/\sqrt{\text{Bo}}\right)},
\nonumber \\
f_{\text{s}}(r) &= \mathrm{K}_1(\sqrt{\beta}r)\quad\text{and}\quad f_{\text{d}}(r)=\sum_{j=1}^4\text{Re}\left[k_j\frac{\mathrm{H}_{-1}(-k_jr)+\mathrm{Y}_1(-k_jr)}{1+\beta/3k_j^2+(4/3)\rmi\epsilon/k_j+(4/3)\epsilon^2k_j}\right].\label{FijID_new}
\end{align}
The equations~\eqref{eq:dimmodel} account for the lateral force and torque balances on each surfer, respectively. The trajectory equations contain a single unknown parameter $F_{\text{p}}$, whose value $F_{\text{p}}=mU/\tau_{\text{v}}$ is directly inferred from the experimentally measured free speed $U$ of a single surfer in isolation. We observe from Table~\ref{tab:param} that $\alpha \ll \xi^2$ for $\gamma/g\geq 1$, the regime in which most of the experiments are conducted~\cite{HoSurfers}, indicating that the dynamic force typically dominates the static force. The dynamic wavefield, which we will plot in \S\ref{Sec:BoundStates}, is obtained by combining Eqs.~\eqref{h1AnalGrav},~\eqref{hFinal},~\eqref{hStatic} and~\eqref{deltaCalc}:
\begin{align}
h(\bs{x},t)&=\frac{mg}{\sigma}\sum_{p=\pm 1}\mu_p\sum_i \mathcal{H}(k_c|\bs{x}-\bs{x}_{i,p}|,t),\nonumber \\
\text{where}\quad\mathcal{H}(r,t)&=\frac{\alpha}{\sqrt{\beta}}\frac{\mathrm{K}_0(\sqrt{\beta}r)}{\text{Bo }\mathrm{K}_0(\sqrt{\text{Bo}})+2\sqrt{\text{Bo}}\,\mathrm{K}_1(\sqrt{\text{Bo}})}+\frac{\xi}{12}\sum_{j=1}^4\text{Re}\left(\frac{C_0(-k_jr)}{1+\beta/3k_j^2+(4/3)\rmi\epsilon/k_j+(4/3)\epsilon^2k_j}\mathrm{e}^{\rmi\omega t}\right).\label{hSurfer}
\end{align}

We proceed by non-dimensionalizing the trajectory equations~\eqref{eq:dimmodel} using $\bs{x}\rightarrow k_c\bs{x}$ and $t\rightarrow tk_cF_{\text{p}}\tau_{\text{v}}/m$:
\begin{align}
\tilde{m}\ddot{\bs{x}}_i+\dot{\bs{x}}_i&=\bs{n}_i+\tilde{F}_{c}\sum_{p,q=\pm 1}\mu_{p}\mu_{q}\sum_{j\neq i}\Phi(|\bs{x}_{j,q}-\bs{x}_{i,p}|)\frac{\bs{x}_{j,q}-\bs{x}_{i,p}}{|\bs{x}_{j,q}-\bs{x}_{i,p}|},\nonumber \\
\tilde{m}\tilde{l}\ddot{\theta}_i+\tilde{l}\dot{\theta}_i&=-\tilde{F}_{c}\sum_{p,q=\pm 1}p\mu_{q}\sum_{j\neq i}\Phi(|\bs{x}_{j,q}-\bs{x}_{i,p}|)\bs{n}_i\times\frac{\bs{x}_{j,q}-\bs{x}_{i,p}}{|\bs{x}_{j,q}-\bs{x}_{i,p}|},\label{NDimEq}
\end{align}
where $\tilde{l}=lk_c$, $\bs{x}_{i,p}=\bs{x}_i-p\mu_{-p}\tilde{l}\bs{n}_i$, and the dimensionless parameters
\begin{align}
\tilde{m} = \frac{k_cF_{\text{p}}\tau_{\text{v}}^2}{m}=k_cU\tau_{\text{v}}\quad \text{and}\quad\tilde{F}_{c} = \frac{F_{c}}{F_{\text{p}}}=\frac{F_{c}\tau_{\text{v}}}{mU}\label{NDimParam}
\end{align}
are defined through the free speed $U=F_{\text{p}}\tau_{\text{v}}/m$ of a single surfer in isolation. Equation~\eqref{NDimEq} is solved using a fourth-order explicit Runge-Kutta method in MATLAB, and the Struve functions in the expression for $\Phi$ are evaluated using the toolbox ``Struve functions" developed by T. P. Theodoulidis. 

\section{Bound states of pairs of surfers}\label{Sec:BoundStates}

For a pair of surfers, Eq.~\eqref{NDimEq} can be written as
\begin{subequations}
\label{NDimEqPair}
\begin{align} 
\tilde{m}\ddot{\bs{x}}_1&=-\dot{\bs{x}}_1+\bs{n}_1+\tilde{F}_c\left\{(\bs{x}_2-\bs{x}_1)\left[\mu_+^2f_{++}+\mu_-^2f_{--}+\mu_+\mu_-(f_{+-}+f_{-+})\right]\right.\nonumber \\
&\phantom{=}\left.+\tilde{l}\mu_+\mu_-\left[(\bs{n}_1-\bs{n}_2)(\mu_+f_{++}-\mu_-f_{--})+\bs{n}_1\left(\mu_-f_{-+}-\mu_+f_{+-}\right)+\bs{n}_2\left(\mu_+f_{-+}-\mu_-f_{+-}\right)\right]\right\},\label{PairA} \\
\tilde{m}\ddot{\bs{x}}_2&=-\dot{\bs{x}}_2+\bs{n}_2-\tilde{F}_c\left\{(\bs{x}_2-\bs{x}_1)\left[\mu_+^2f_{++}+\mu_-^2f_{--}+\mu_+\mu_-(f_{+-}+f_{-+})\right]\right.\nonumber \\
&\phantom{=}\left.+\tilde{l}\mu_+\mu_-\left[(\bs{n}_1-\bs{n}_2)(\mu_+f_{++}-\mu_-f_{--})+\bs{n}_1\left(\mu_-f_{-+}-\mu_+f_{+-}\right)+\bs{n}_2\left(\mu_+f_{-+}-\mu_-f_{+-}\right)\right]\right\},\label{PairB}\\
\tilde{m}\tilde{l}\ddot{\theta}_1&=-\tilde{l}\dot{\theta}_1+\tilde{F}_c\bs{n}_1\times\left\{(\bs{x}_2-\bs{x}_1)\left[\mu_+(f_{+-}-f_{++})-\mu_-(f_{-+}-f_{--})\right]+\mu_+\mu_-\tilde{l}\left(f_{++}+f_{--}-f_{+-}-f_{-+}\right)\bs{n}_2\right\},\label{PairC} \\
\tilde{m}\tilde{l}\ddot{\theta}_2&=-\tilde{l}\dot{\theta}_2-\tilde{F}_c\bs{n}_2\times\left\{(\bs{x}_2-\bs{x}_1)\left[\mu_+(f_{-+}-f_{++})-\mu_-(f_{+-}-f_{--})\right]-\mu_+\mu_-\tilde{l}\left(f_{++}+f_{--}-f_{+-}-f_{-+}\right)\bs{n}_1\right\},\label{PairD}
\end{align}
\end{subequations}
where $f_{pq}=\Phi(|\bs{\delta}_{pq}|)/|\bs{\delta}_{pq}|$ and $\bs{\delta}_{pq}=\bs{x}_{2,p}-\bs{x}_{1,q}$; specifically,
\begin{align}
\bs{\delta}_{++}&=\bs{x}_{2}-\bs{x}_{1}-\mu_-\tilde{l}\left(\bs{n}_2-\bs{n}_1\right),\quad
\bs{\delta}_{--}=\bs{x}_{2}-\bs{x}_{1}+\mu_+\tilde{l}\left(\bs{n}_2-\bs{n}_1\right),\quad
\bs{\delta}_{+-}=\bs{x}_{2}-\bs{x}_{1}-\tilde{l}\left(\mu_-\bs{n}_2+\mu_+\bs{n}_1\right),\nonumber \\
\text{and}\quad \bs{\delta}_{-+}&=\bs{x}_{2}-\bs{x}_{1}+\tilde{l}\left(\mu_+\bs{n}_2+\mu_-\bs{n}_1\right).\label{deltapm}
\end{align} 

\begin{figure*}[ht]
  \centering
    \includegraphics[width=1\textwidth]{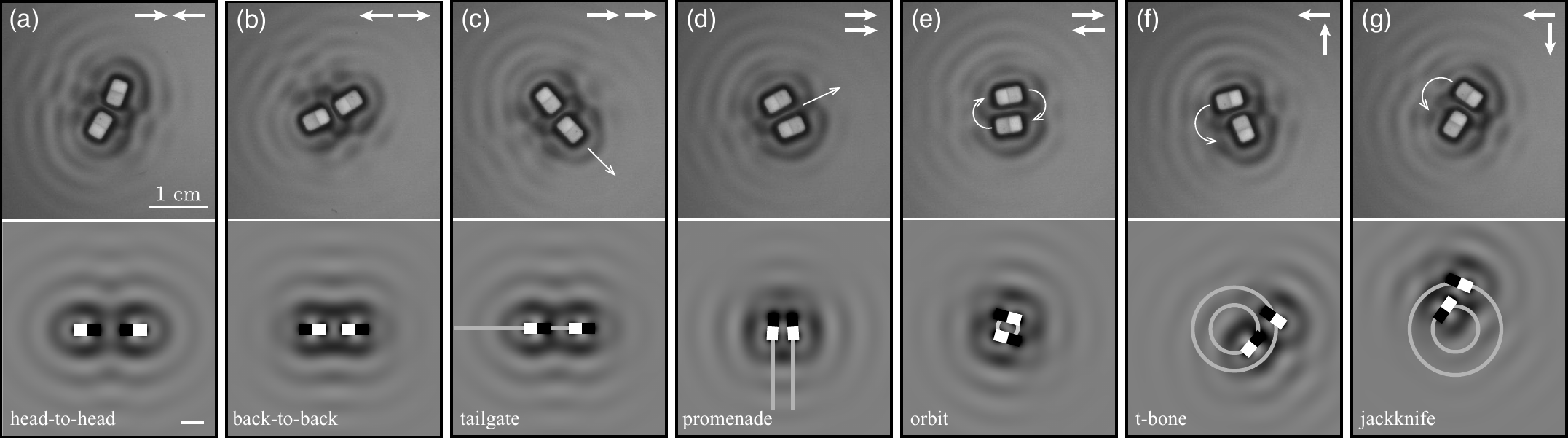}
      \caption{Bound states of pairs of surfers, obtained in experiment~\cite{HoSurfers} (top row) and numerical simulations of~\eqref{NDimEqPair} with different initial conditions (bottom row). (a) Head-to-head, (b) back-to-back, (c) tailgate, (d) promenade, (e) orbit, (f) t-bone and (g) jackknife. The forcing frequency is $f = 100$ Hz and forcing acceleration is $\gamma/g=3.3$, for which the surfer free speed is $U = 1.9$ mm/s. The values of the parameters are given in Table~\ref{tab:param}. The associated wavefields are given by Eq.~\eqref{hSurfer} evaluated at $t=0$. Scale bar in numerical simulations denotes the capillary wavelength $\lambda_c$.}
        \label{BoundStates}
\end{figure*}

Numerical simulations of Eq.~\eqref{NDimEqPair} demonstrate that our model recovers the seven different interaction modes exhibited by two surfers of equal size and speed [Fig.~\ref{BoundStates}]. 
In the head-to-head mode [Fig.~\ref{BoundStates}(a)] the two surfer bows face each other, while in the back-to-back mode [Fig.~\ref{BoundStates}(b)] the two surfer sterns face each other. While these modes are static, the remaining five modes are dynamic. In the tailgating mode [Fig.~\ref{BoundStates}(c), Supplemental Video 2], the surfers are aligned along their major axis, with the bow of one surfer pointing toward the stern of the other, and they move with constant speed along a rectilinear trajectory. In the promenade mode [Fig.~\ref{BoundStates}(d), Supplemental Video 3], they proceed side by side with constant speed along a rectilinear trajectory.  In the orbiting mode [Fig.~\ref{BoundStates}(e), Supplemental Video 4], the two surfers orbit around the system’s fixed center of mass. In the t-bone mode [Fig.~\ref{BoundStates}(f), Supplemental Video 5], the two major axes are perpendicular to each other and the bow of one surfer points toward the stern of the other, while they both execute a circular trajectory. The jackknife mode [Fig.~\ref{BoundStates}(g), Supplemental Video 5] has a similar configuration except the stern of one surfer points toward the stern of the other. 

We proceed by considering the existence and stability of the bound states shown in Fig.~\ref{BoundStates}, as predicted by the model~\eqref{NDimEqPair}. In Appendices~\ref{App:Stability_Rect} and~\ref{App:Stability_Rot}, respectively, we derive a framework for assessing the linear stability of rectilinear (head-to-head, back-to-back, tailgating, promenading) and rotating (orbiting, t-bone, jackknife) states. We then apply our framework to understand, in turn, one-dimensional rectilinear modes (head-to-head, back-to-back, tailgating) in \S\ref{SSec:1DOF}, the promenade mode in \S\ref{SSec:Prom}, and the rotating modes in \S\ref{SSec:Rot}. To accomplish this, we rewrite Eq.~\eqref{NDimEqPair} in terms of the variables
\begin{align}
\bs{\sigma}=\bs{x}_1+\bs{x}_2\quad\text{and}\quad \bs{\delta}=\bs{x}_2-\bs{x}_1.\label{SigmaDeltaDef}
\end{align}
Adding and subtracting Eqs.~\eqref{PairA}-\eqref{PairB} and Eqs.~\eqref{PairC}-\eqref{PairD}, we obtain
\begin{subequations}
\label{TwoSurfCM2}
\begin{align}
m\ddot{\bs{\sigma}}&=-\dot{\bs{\sigma}}+\bs{n}_1+\bs{n}_2,\label{CMA} \\
m\ddot{\bs{\delta}}&=-\dot{\bs{\delta}}+\bs{n}_2-\bs{n}_1-2\tilde{F}_c\left[\bs{\delta}\mathcal{F}+\tilde{l}\mu_+\mu_-(\bs{n}_1\mathcal{T}_1-\bs{n}_2\mathcal{T}_2)\right],\label{CMB} \\
m\tilde{l}(\ddot{\theta}_1+\ddot{\theta}_2)&=-\tilde{l}(\dot{\theta}_1+\dot{\theta}_2)+\tilde{F}_c\bs{\delta}\times(\bs{n}_1\mathcal{T}_1-\bs{n}_2\mathcal{T}_2),\label{CMC} \\
m\tilde{l}(\ddot{\theta}_1-\ddot{\theta}_2)&=-\tilde{l}(\dot{\theta}_1-\dot{\theta}_2)+\tilde{F}_c\left[\bs{\delta}\times\left(\bs{n}_1\mathcal{T}_1+\bs{n}_2\mathcal{T}_2\right)+2\mu_+\mu_-\tilde{l}\mathcal{S}\bs{n}_1\times\bs{n}_2\right],\label{CMD}
\end{align}
\end{subequations}
where
\begin{align}
\mathcal{F} &= \mu_+^2f_{++}+\mu_-^2f_{--}+\mu_+\mu_-(f_{+-}+f_{-+}),\quad \mathcal{S} = f_{++}+f_{--}-f_{-+}-f_{+-}\nonumber \\
\mathcal{T}_1 &= \mu_+f_{++}-\mu_+f_{+-}+\mu_-f_{-+}-\mu_-f_{--},\quad
 \mathcal{T}_2=\mu_+f_{++}+\mu_-f_{+-}-\mu_+f_{-+}-\mu_-f_{--}.
\end{align}

\subsection{Head-to-head, back-to-back and tailgating modes}\label{SSec:1DOF}

The head-to-head mode [Fig.~\ref{BoundStates}(a)] centered at the origin and oriented along the $x$-axis is given by $\bs{\sigma}=\boldsymbol{0}$, $\bs{\delta}=( d,0)$ (corresponding to $\bs{x}_1=( -d/2,0)$, $\bs{x}_2=( d/2,0)$), $\bs{n}_1=( 1,0)$ and $\bs{n}_2=( -1,0)$, where $d$ is the distance between the centers of mass. Substituting this solution into Eq.~\eqref{TwoSurfCM2} and defining $\mu_0$ through  $\mu_{\pm}=1/2\pm\mu_0$, we obtain a single algebraic equation that determines $d$:
\begin{align}
1&=\tilde{F}_cF_{\text{H}}(d),\quad\text{where}\quad F_{\text{H}}(d)=-d\left(\mu_+^2f_{++}+\mu_-^2f_{--}+2\mu_+\mu_-f_{+-}\right)-2\tilde{l}\mu_+\mu_-\left[\mu_+f_{++}-\mu_-f_{--}-2\mu_0f_{+-}\right]\nonumber \\
\text{and}&\quad \left|\bs{\delta}_{++}\right|=d+2\mu_-\tilde{l},\quad  \left|\bs{\delta}_{--}\right|=d-2\mu_+\tilde{l},\quad  \left|\bs{\delta}_{-+}\right|= \left|\bs{\delta}_{+-}\right|=d-2\tilde{l}\mu_0.\label{FrontToFront}
\end{align}
Similarly, the back-to-back mode [Fig.~\ref{BoundStates}(b)] is given by $\bs{\sigma}=\boldsymbol{0}$, $\bs{\delta}=( d,0)$, $\bs{n}_1=( -1,0)$ and $\bs{n}_2=( 1,0)$, from which we obtain
\begin{align}
-1&=\tilde{F}_cF_{\text{B}}(d),\quad\text{where}\quad F_{\text{B}}(d)=-d\left(\mu_+^2f_{++}+\mu_-^2f_{--}+2\mu_+\mu_-f_{+-}\right)+2\tilde{l}\mu_+\mu_-\left[\mu_+f_{++}-\mu_-f_{--}-2\mu_0f_{+-}\right]\nonumber \\
\text{and}&\quad \left|\bs{\delta}_{++}\right|=d-2\mu_-\tilde{l},\quad \left|\bs{\delta}_{--}\right|=d+2\mu_+\tilde{l},\quad \left|\bs{\delta}_{-+}\right|=\left|\bs{\delta}_{+-}\right|=d+2\tilde{l}\mu_0.\label{BackToBack}
\end{align}
The tailgating mode [Fig.~\ref{BoundStates}(c)] with speed $v$ is given by $\bs{\sigma}= ( vt,0)$, $\bs{\delta}=( d,0)$ (corresponding to $\bs{x}_1=( -d/2+vt,0)$, $\bs{x}_2=( d/2+vt,0)$), and $\bs{n}_1=\bs{n}_2=( 1,0)$. The first equation in Eq.~\eqref{TwoSurfCM2} implies that $v=1$, while the second reduces to
\begin{align}
F_{\text{T}}(d)=0,\quad\text{where}&\quad F_{\text{T}}(d)=-d\left[(\mu_+^2+\mu_-^2)f_{++}+\mu_+\mu_-(f_{+-}+f_{-+})\right]-\tilde{l}\mu_+\mu_-(f_{-+}-f_{+-})\nonumber \\
\text{and}&\quad \left|\bs{\delta}_{++}\right|=\left|\bs{\delta}_{--}\right|=d,\quad\left|\bs{\delta}_{+-}\right|=d-\tilde{l},\quad \left|\bs{\delta}_{-+}\right|=d+\tilde{l}.\label{Following}
\end{align}

 \begin{figure}
\begin{center}
\includegraphics[width=1\textwidth]{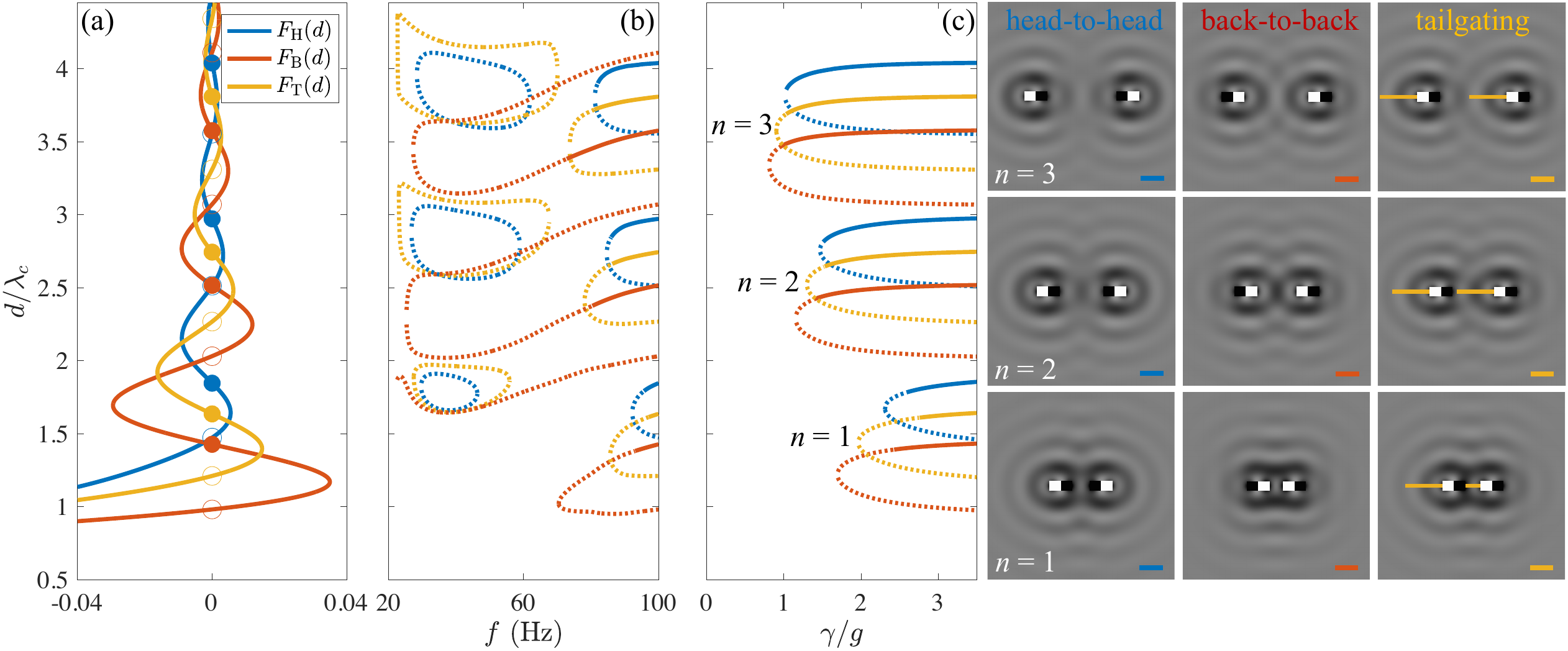}
 \end{center}
 \caption{(a) Force curves corresponding to the head-to-head (blue, Eq.~\eqref{FrontToFront}), back-to-back (red, Eq.~\eqref{BackToBack}) and tailgating (yellow, Eq.~\eqref{Following}) modes, for $f = 100$ Hz and $\gamma/g=3.3$. Filled (unfilled) circles correspond to stable (unstable) solutions. (b) Dependence of the distance $d$ on the forcing frequency $f$, as described in \S\ref{SSec:1DOF}, for surfer pairs in each of the three modes. Solid (dashed) curves correspond to stable (unstable) solutions. (c) The dependence of the distance $d$ between surfers on the forcing acceleration $\gamma$ for fixed forcing frequency $f = 100$ Hz.  The three rightmost columns show, for the mode order $n$ indicated, the (unique) stable mode for $\gamma/g=3.3$ and $f = 100$ Hz. The corresponding wavefield is computed using Eq.~\eqref{hSurfer} evaluated at $t=0$, and scale bars denote the capillary wavelength $\lambda_c$. Movies of the tailgating modes are shown in Supplemental Video 2.
}
 \label{Fig:ForceCurves}
 \end{figure}

The force curves $F_{\text{H}}(d)$, $F_{\text{B}}(d)$ and $F_{\text{T}}(d)$ are shown in Fig.~\ref{Fig:ForceCurves}(a). The equilibrium distances $d$ are found numerically using bisection; since $\tilde{F}_c\gg 1$ in the parameter regime of interest (Table~\ref{tab:param}), the equilibrium distances are well-approximated by the roots of the functions $F_{\text{H}}$, $F_{\text{B}}$ and $F_{\text{T}}$. The stability of the equilibria is assessed using the framework detailed in Appendix~\ref{App:Stability_Rect}. The dependence of $d$ on the forcing frequency $f$ is shown in Fig.~\ref{Fig:ForceCurves}(b). As in the experiments (see Fig.~3(f) in~\cite{HoSurfers}), both $f$ and $\gamma$ (and thus $\xi$) are varied together; specifically, $\gamma/g$ increases from 1.1 to 3.3 as $f$ is varied from 50 to 100 Hz. 
The dimensionless parameters $\tilde{m}$ and $\tilde{l}$ depend on the surfer free speed $U$, which in turn varies with both $\gamma$ and $f$. The values of $U$ and $\xi$ are thus inferred from the experimental data in Supplementary Fig. S1 of~\cite{HoSurfers} using linear interpolation or extrapolation. The dependence of $d$ on the forcing acceleration $\gamma$ for a fixed forcing frequency $f=100$ Hz is shown in Fig.~\ref{Fig:ForceCurves}(c). 

From Fig.~\ref{Fig:ForceCurves}(b-c), we observe that, for each of the three modes considered, there is a quantized set of stable solutions (solid lines) separated by unstable ones (dashed lines). Specifically, in the stable head-to-head, back-to-back and tailgating modes, the centers-of-mass are separated by roughly integer multiples of the capillary wavelength: $d=n\lambda_c$, $d=(n-1/2)\lambda_c$ and $d=(n-1/4)\lambda_c$, respectively, where $n\geq 2$. From Fig.~\ref{Fig:ForceCurves}(b) [Fig.~\ref{Fig:ForceCurves}(c)], we observe that stable modes exist over a larger range of $f$ ($\gamma$) values as $n$ increases. We also note that, as shown in Fig.~\ref{Fig:ForceCurves}(b), there are unstable families of solutions at relatively low frequencies ($f < 70$ Hz). The head-to-head and back-to-back equilibria will play a role in the next section (\S\ref{SSec:Prom}), where we discuss the so-called promenade mode.

\subsection{Promenade mode}\label{SSec:Prom}

The promenade mode [Fig.~\ref{BoundStates}(d)], in which surfers move side-by-side at a constant velocity orthogonal to the line connecting their centers, is given by $\bs{\sigma}=( 0,vt)$, $\bs{\delta}=( d,0)$ (corresponding to $\bs{x}_1(t)=( -d/2,vt)$, $\bs{x}_2(t)=( d/2,vt)$), $\theta_1(t)=\pi-\varphi_2$ 
and $\theta_2(t)=\varphi_2$. 
Substituting this solution into Eq.~\eqref{TwoSurfCM2} we obtain a system of equations that determines the distance $d$ between surfers, their speed $v$ and orientation $\varphi_2$: 
\small
\begin{align}
v&=\sin\varphi_2,\nonumber \\
0&=F_{\text{P}}(d,\varphi_2)\equiv-\cos\varphi_2+\tilde{F}_c\left\{\left(\mu_+^2f_{++}+\mu_-^2f_{--}+2\mu_+\mu_-f_{+-}\right)d-2\tilde{l}\mu_+\mu_-\cos\varphi_2\left[\mu_+f_{++}-\mu_-f_{--}-2\mu_0f_{+-}\right]\right\},\nonumber \\
0&=T_{\text{P}}(d,\varphi_2)\equiv d\left[\mu_+f_{++}-\mu_-f_{--}-2\mu_0f_{+-}\right]-2\mu_+\mu_-\tilde{l}\cos\varphi_2\left(f_{++}+f_{--}-2f_{+-}\right),\nonumber \\
\text{where}&\quad \left|\bs{\delta}_{++}\right|=\left|d-2\mu_-\tilde{l}\cos\varphi_2\right|,\quad \left|\bs{\delta}_{--}\right|=\left|d+2\mu_+\tilde{l}\cos\varphi_2\right|,\nonumber \\
\text{and}&\quad\left|\bs{\delta}_{+-}\right|^2=\left|\bs{\delta}_{-+}\right|^2=d^2+\tilde{l}^2\left(\mu_+^2+\mu_-^2-2\mu_+\mu_-\cos 2\varphi_2\right)+4d\tilde{l}\mu_0\cos\varphi_2.\label{PromEqs}
\end{align}
\normalsize
The first (second) equation in Eq.~\eqref{PromEqs} represents the force balance in the transverse (lateral) direction, while the third equation represents the torque balance. The distinct promenade modes are found numerically by finding the roots of $F_{\text{P}}(d,\varphi_2)$ 
and $T_{\text{P}}(d,\varphi_2)$, 
which constitutes a system of two equations in two unknowns. Specifically, we compute the zero contours of the two functions using MATLAB and locate their intersections~\cite{Schwarz_Intersections}, as depicted in Appendix Fig.~\ref{PromContour}. The stability of the solutions is assessed using the framework described in Appendix~\ref{App:Stability_Rect}. We assume that $v \geq 0$ and thus restrict our attention to $0\leq\varphi_2\leq\pi$.

\begin{figure*}[ht]
  \centering
    \includegraphics[width=1\textwidth]{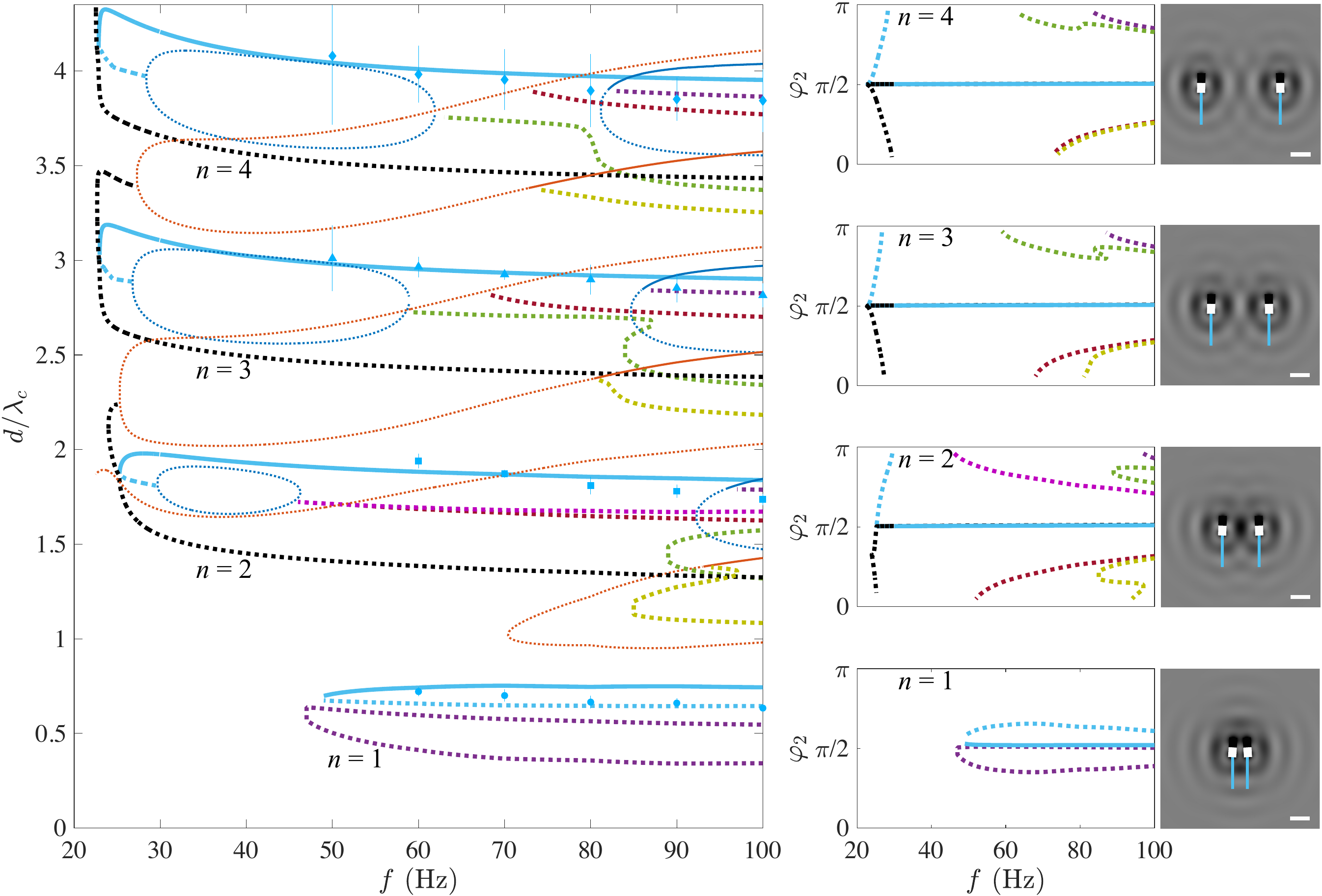}
      \caption{The dependence of the promenade mode equilibria on the forcing frequency $f$, as obtained by solving Eq.~\eqref{PromEqs} using the procedure described in \S\ref{SSec:Prom}. The large panel shows the dependence on $f$ of the distance $d$ between the surfers' centers of mass. Stable (unstable) promenade mode solutions are indicated by the solid (dashed) curves. Data points correspond to the values of $d-w$ obtained in experiments (see Fig.~3f in~\cite{HoSurfers}), the $w$--term accounting for the surfers' finite width. In the experiments,  $f$ ranges from 50-100 Hz in increments of
10 Hz, and the corresponding values of $\gamma/g$ are 1.1, 1.5, 2.0, 2.3,
3.0 and 3.3. The head-to-head (blue) and back-to-back (red) modes from Fig.~\ref{Fig:ForceCurves}(b) are superimposed.  The middle column shows 
the corresponding orientation angle $\varphi_2$ 
for the mode order $n$ indicated. For a given mode order, the colors correspond to those in the large panel. The rightmost column shows, for each $n$, the (unique) stable promenade mode for the combination $f = 100$ Hz and $\gamma/g=3.3$, and the corresponding wavefield~\eqref{hSurfer} evaluated at $t=0$. Scale bars denote the capillary wavelength $\lambda_c$. Movies of these promenade modes are shown in Supplemental Video 3.
      }
        \label{PromPlotsFreq}
\end{figure*}

The dependence of the equilibrium distance $d$ and orientation angle $\varphi_2$ 
on the forcing frequency $f$ is shown in Fig.~\ref{PromPlotsFreq}. As in Fig.~\ref{Fig:ForceCurves}(b), both $f$ and $\gamma$ are varied together, with the intermediate values extrapolated from the experimental data as detailed in \S\ref{SSec:1DOF} and the caption of Fig.~\ref{PromPlotsFreq}. We observe that the stable (solid curves) promenade modes are roughly quantized on the capillary wavelength, with separation distance  $d\approx n\lambda_c$ for $n\in\mathbb{N}$. The stable states have angle $\varphi_2\gtrsim\pi/2$, 
indicating that the surfers are approximately oriented along their direction of motion. The model also predicts a number of unstable (dashed curves) solutions with a variety of separation distances and angles. The stable equilibrium distances exhibit excellent agreement with experiment, with the theory correctly capturing the slight decrease in $d/\lambda_c$ with increasing $f$. The agreement between theory and experiment improves at lower values of $f$, presumably because the quasipotential approximation for the wavefield in Eq.~\eqref{ViscGravPotFlow} is valid for $\epsilon\ll 1$, and the reciprocal Reynolds number scales as $\epsilon \sim \omega^{1/3}$ from Eq.~\eqref{ParamDef1}.  We note that, for the experimental data points in Fig.~\ref{PromPlotsFreq}, the surfer width $w$ is subtracted from $d$. This correction accounts for the fact that, while a surfer is represented as a pair of point sources in the model, experimental observations indicate that a surfer generates waves along its whole perimeter. We also note that, as $f$ is varied for $n=$ 2, 3 and 4, most of the solution branches bifurcate into the head-to-head (blue) and back-to-back (red) modes as $\varphi_2\rightarrow \pi^-$ ($\varphi_2\rightarrow 0^+$).

Since $v=\sin\varphi_2$ and $\varphi_2\gtrsim\pi/2$, 
the theory predicts that the promenade speed is just slightly less than the free speed of a single surfer.  In experiments the promenade speed is also always less than the free speed, but can go down to as low as 50\% of the free speed. A similar quantitative discrepancy was observed in a study on oil droplets that bounce on the surface of a vertically vibrating fluid bath, pairs of which also executed the promenade mode~\cite{Arbelaiz2018}. In that study, the discrepancy was resolved by modeling the coupling between the droplets' horizontal and vertical dynamics. While the surfer model presented herein neglects the vertical dynamics entirely, presumably an analogous extension of the model would lead to predicted promenade speeds that are closer to those observed in experiments.

\begin{figure*}[ht]
  \centering
    \includegraphics[width=.75\textwidth]{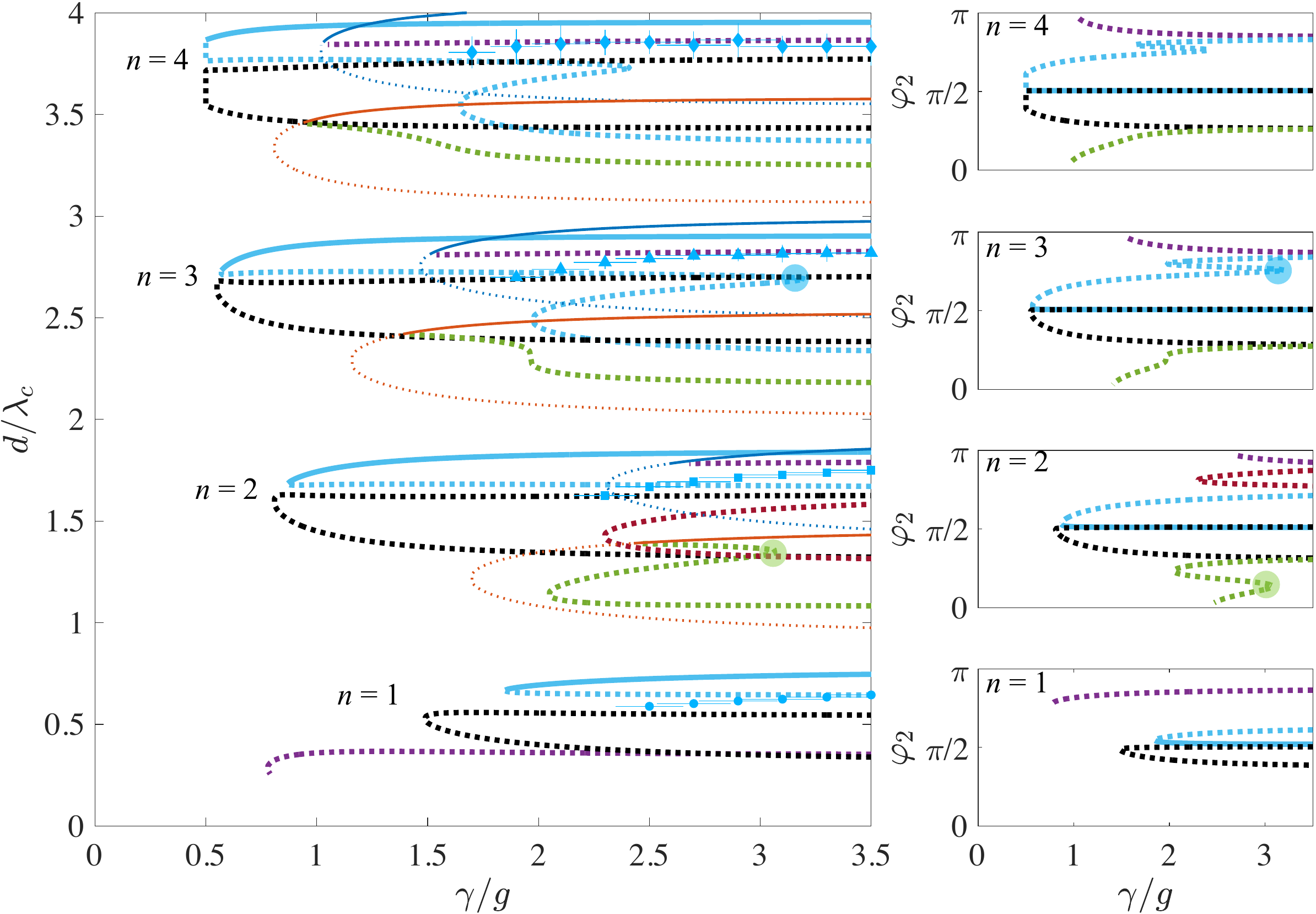}
      \caption{The dependence of the promenade mode equilibria on the forcing acceleration $\gamma$, for the fixed forcing frequency $f = 100$ Hz. The head-to-head (blue) and back-to-back (red) modes from Fig.~\ref{Fig:ForceCurves}(c) are superimposed. Data points correspond to the values of $d-w$ obtained in experiments (see Fig.~3e in~\cite{HoSurfers}). See the caption of Fig.~\ref{PromPlotsFreq} for more details.             }
        \label{PromPlotsGm}
\end{figure*}

Figure~\ref{PromPlotsGm} shows the dependence of the equilibrium distance $d$ and orientation angle $\varphi_2$ 
on the forcing acceleration $\gamma$ for the forcing frequency $f = 100$ Hz, the largest value of $f$ considered in experiments.  The predicted equilibrium distances $d$ exhibit adequate agreement with experiment, and correctly capture a number of trends: namely, that $d$ is quantized on the capillary wavelength $\lambda_c$ and remains slightly below integer multiples of $\lambda_c$; $d$ increases very slightly with forcing acceleration $\gamma$; and, that the critical $\gamma$ above which stable promenade solutions appear decreases with the mode order $n$, as was the case with the head-to-head, back-to-back and tailgating modes [Fig.~\ref{Fig:ForceCurves}(c)]. However, the theoretically predicted values of $d$ are systematically larger than those obtained in experiment, presumably due to the fact that the quasipotential approximation is less accurate at larger frequencies. We note that, for $\gamma\approx 3g$, the theory predicts two small regions of ``exotic" promenade modes, highlighted by the green and blue circles, where $d/\lambda_c\approx 1.25$ and 2.75 and $\varphi_2\approx \pi/8$ and $3\pi/4$, respectively. These exotic states coexist with the other promenade modes, and presumably exist in a corner of parameter space too small to be accessed by experiments.

\subsection{Orbiting, jackknife and t-bone modes}\label{SSec:Rot}

Circular orbit solutions, in which two surfers traverse a circular orbit with constant angular frequency $\omega_0$, are given by $\bs{\delta}=d( \cos\omega_0 t,\sin\omega_0 t)$, $\bs{n}_1=( \cos(\omega_0 t+\varphi_1),\sin(\omega_0 t+\varphi_1))$ and $\bs{n}_2 =( \cos(\omega_0 t+\varphi_2),\sin(\omega_0 t+\varphi_2))$. We substitute this solution into Eq.~\eqref{TwoSurfCM2}. To simplify the resulting system of equations, we take the cross product of $\bs{\delta}$ with Eq.~\eqref{CMB}, and add to it the product of Eq.~\eqref{CMC} and $2\mu_+\mu_-\tilde{l}$:
\begin{align}
\omega_0\left(d^2+4\mu_+\mu_-\tilde{l}^2\right)=d(\sin\varphi_2-\sin\varphi_1)\quad\Rightarrow\quad \omega_0 = \frac{d(\sin\varphi_2-\sin\varphi_1)}{d^2+4\mu_+\mu_-\tilde{l}^2}.\label{BlueBox2}
\end{align}
We then take the cross product of Eq.~\eqref{CMB} with $ (\bs{n}_1\mathcal{T}_1-\bs{n}_2\mathcal{T}_2)$, and add to it the product of Eq.~\eqref{CMC} and $2\mathcal{F}$: 
\begin{align}
-\tilde{m}\omega_0^2d\left(\mathcal{T}_1\sin\varphi_1-\mathcal{T}_2\sin\varphi_2\right)-d\omega_0\left(\mathcal{T}_1\cos\varphi_1-\mathcal{T}_2\cos\varphi_2\right)+(\mathcal{T}_1-\mathcal{T}_2)\sin(\varphi_2-\varphi_1)=-4\tilde{l}\omega_0\mathcal{F}.\label{GreenBox2}
\end{align}
Equation~\eqref{CMC} reduces to
\begin{align}
2\omega_0\tilde{l}=\tilde{F}_cd\left(\mathcal{T}_1\sin\varphi_1-\mathcal{T}_2\sin\varphi_2\right),
\label{TorqueSum1}
\end{align}
while Eq.~\eqref{CMD} reduces to
\begin{align}
d\left(\mathcal{T}_1\sin\varphi_1+\mathcal{T}_2\sin\varphi_2\right)+2\mu_+\mu_-\tilde{l}\mathcal{S}\sin(\varphi_2-\varphi_1)=0.\label{RedBox}
\end{align}
Using Eq.~\eqref{deltapm}, $\mathcal{F}$, $\mathcal{S}$, $\mathcal{T}_1$ and $\mathcal{T}_2$ are evaluated using the formulas
\begin{align}
\left|\bs{\delta}_{++}\right|^2&=d^2+\left(2\tilde{l}\mu_-\sin\frac{\varphi_2-\varphi_1}{2}\right)^2-2d\tilde{l}\mu_-\left(\cos\varphi_2-\cos\varphi_1\right),\nonumber \\
\left|\bs{\delta}_{--}\right|^2&=d^2+\left(2\tilde{l}\mu_+\sin\frac{\varphi_2-\varphi_1}{2}\right)^2+2d\tilde{l}\mu_+\left(\cos\varphi_2-\cos\varphi_1\right),\nonumber \\
\left|\bs{\delta}_{+-}\right|^2&=d^2+\tilde{l}^2\left[\mu_+^2+\mu_-^2+2\mu_+\mu_-\cos(\varphi_2-\varphi_1)\right]-2d\tilde{l}\left(\mu_+\cos\varphi_1+\mu_-\cos\varphi_2\right),\nonumber \\
\left|\bs{\delta}_{-+}\right|^2&=d^2+\tilde{l}^2\left[\mu_+^2+\mu_-^2+2\mu_+\mu_-\cos(\varphi_2-\varphi_1)\right]+2d\tilde{l}\left(\mu_-\cos\varphi_1+\mu_+\cos\varphi_2\right).
\end{align}
Using Eq.~\eqref{BlueBox2} to eliminate $\omega_0$, the system of three equations~\eqref{GreenBox2}-\eqref{RedBox} thus defines the three unknowns $d$, $\varphi_1$ and $\varphi_2$. The stability of circular orbits is assessed using the framework described in Appendix~\ref{App:Stability_Rot}.

The orbiting mode [Fig.~\ref{BoundStates}(e)] is a special case in which the surfers orbit their fixed center of mass while remaining diametrically opposed to each other. Equation~\eqref{RedBox} is trivial in this mode, since $\varphi_2-\varphi_1 =\pi$ and thus $\left|\bs{\delta}_{+-}\right|=\left|\bs{\delta}_{-+}\right|$, so $\mathcal{T}_1=\mathcal{T}_2$. After using Eq.~\eqref{BlueBox2}, Eqs.~\eqref{GreenBox2} and~\eqref{TorqueSum1} comprise a system of two equations in the two unknowns $d$ and $\varphi_2$, which may be solved using the method described in \S\ref{SSec:Prom}. We assume that the surfers orbit in the counterclockwise sense ($\omega_0 > 0$), and thus restrict our attention to $0\leq\varphi_2\leq\pi$. 

\begin{figure*}[ht]
  \centering
    \includegraphics[width=1\textwidth]{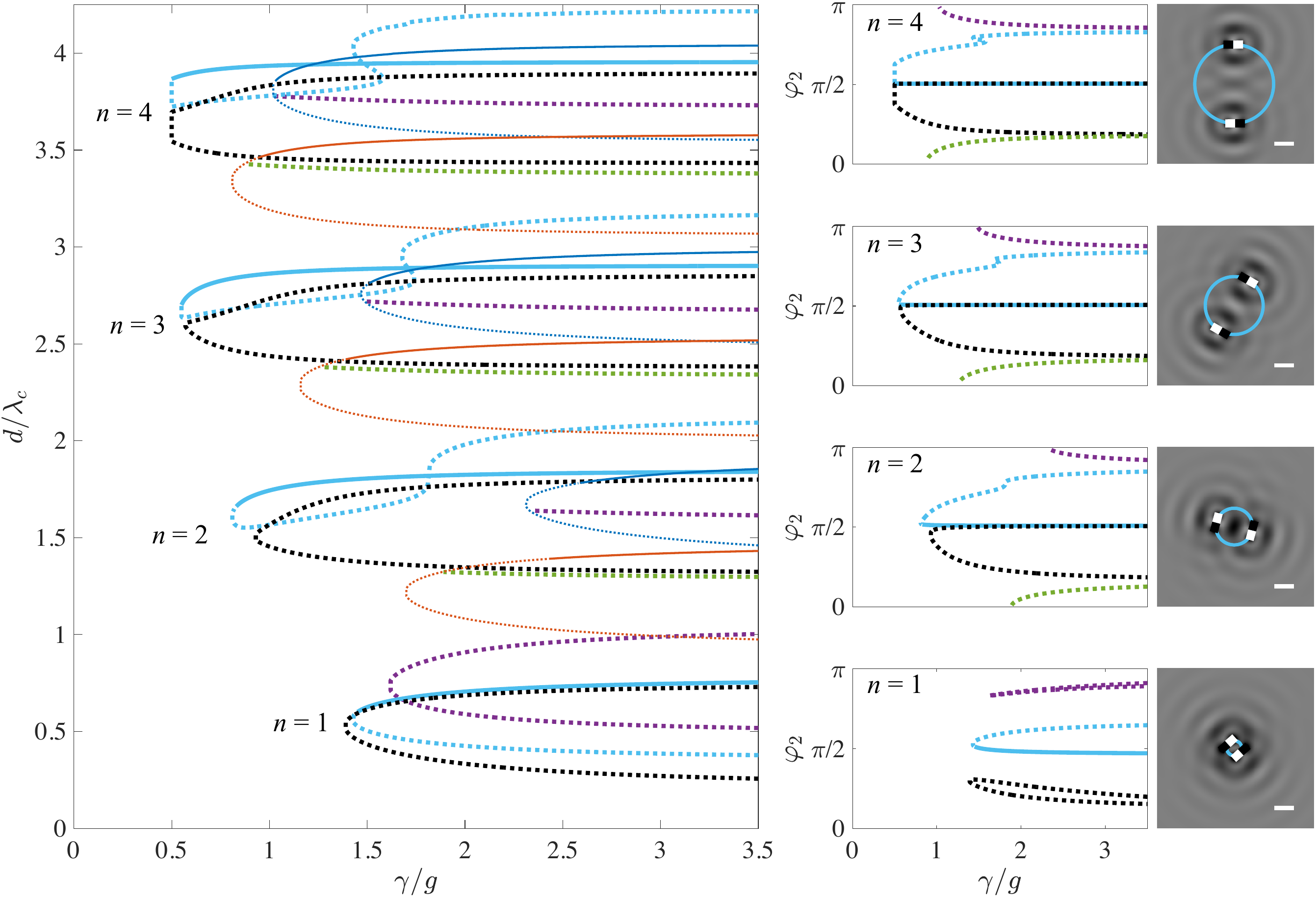}
      \caption{Orbiting modes of surfer pairs, obtained by solving Eqs.~\eqref{GreenBox2} and~\eqref{TorqueSum1} for the orbital diameter $d$ and orientation angle $\varphi_2=\varphi_1+\pi$. The large panel shows the dependence of $d$ on the forcing acceleration $\gamma$. Stable (unstable) orbiting modes are indicated by the solid (dashed) curves. The head-to-head (blue) and back-to-back (red) modes from Fig.~\ref{Fig:ForceCurves}(b) are superimposed. The middle column shows $\varphi_2$ for the mode order indicated.  For a given mode order $n$, curves of the same color indicate the same solution branch. The rightmost column shows, for each $n$, the (unique) stable orbiting mode for $\gamma/g=3.3$ and the corresponding wavefield~\eqref{hSurfer} evaluated at $t=0$. Scale bars denote the capillary wavelength $\lambda_c$. Movies of these orbiting modes are shown in Supplemental Video 4.}
        \label{OrbPlot}
\end{figure*}

The dependence of $d$ and $\varphi_2$ on the forcing acceleration $\gamma$ is shown in Fig.~\ref{OrbPlot}. As with the promenade mode [Fig.~\ref{PromPlotsGm}], we observe that the stable (solid curves) orbiting modes are roughly quantized on the capillary wavelength, with separation distance  $d\approx n\lambda_c$ for $n\in\mathbb{N}$. The stable states have angle $\varphi_2\gtrsim \pi/2$, indicating that the surfers remain roughly tangent to the circle they traverse. As with the promenade mode, for $n=$ 2, 3 and 4, two branches of unstable orbiting modes bifurcate into the head-to-head (blue) and back-to-back (red) modes as $\varphi_2\rightarrow \pi^-$ ($\varphi_2\rightarrow 0^+$). Using Eq.~\eqref{BlueBox2}, we deduce that the orbital speed $d\omega_0/2$ increases with orbit order $n$ and remains less than unity, the free speed of a single surfer. While the orbital speeds in the $n=1$ and $n=2$ modes, respectively, are predicted to be roughly 60\% and 90\% of the free speed, in experiments they are roughly equal to the free speed (see Supplementary Table S1 in~\cite{HoSurfers}).

\begin{figure*}[ht]
  \centering
    \includegraphics[width=1\textwidth]{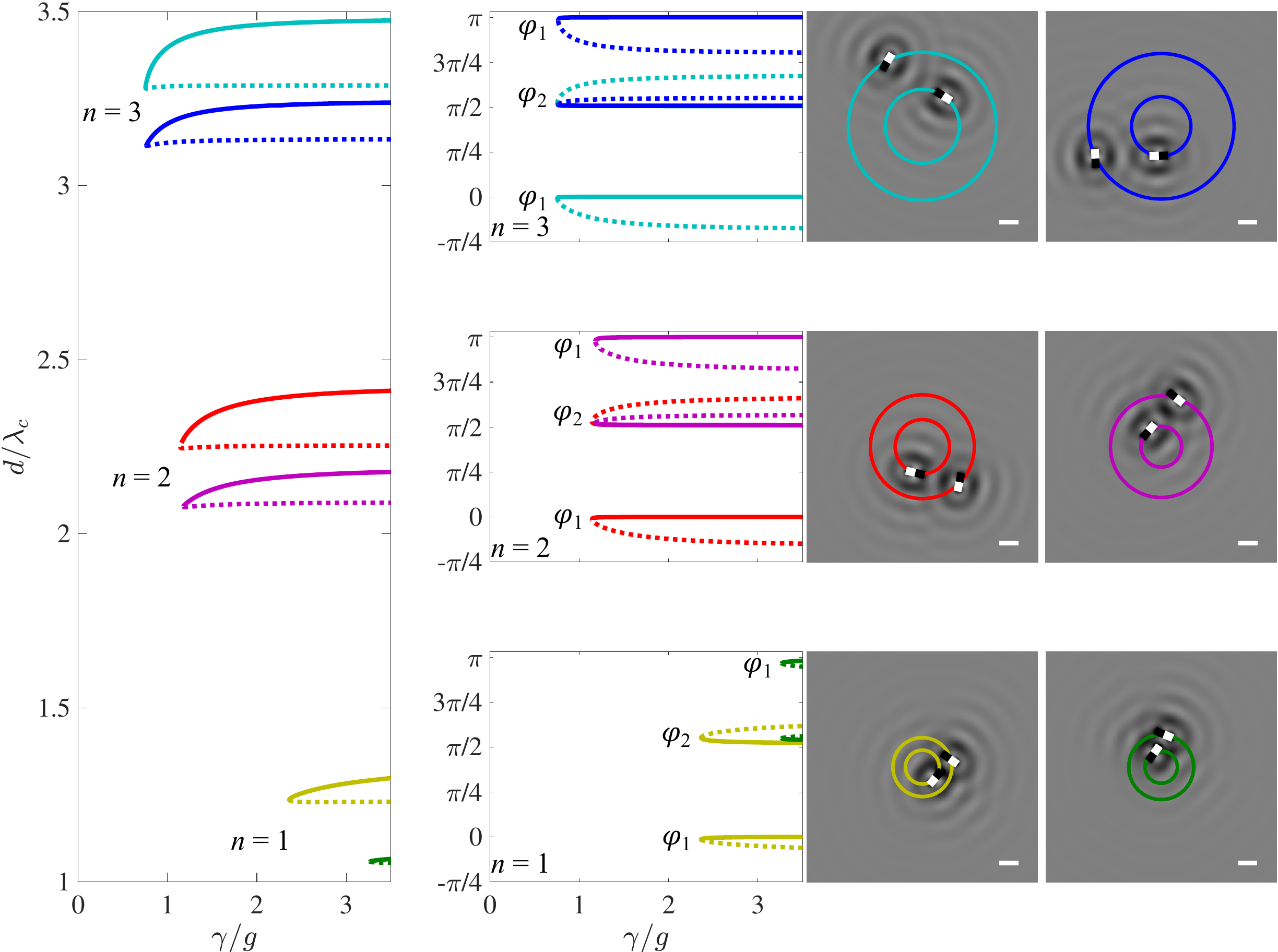}
      \caption{T-bone (yellow, red, cyan) and jackknife (green, magenta, blue) modes, as obtained by solving Eqs.~\eqref{GreenBox2}-\eqref{RedBox} for the forcing frequency $f = 100$ Hz. Stable (unstable) states are indicated by the solid (dashed) lines. The left column shows the dependence of the distance $d$ between surfers on the forcing acceleration $\gamma$. The panels in the second column show the corresponding orientation angles $\varphi_1$ and $\varphi_2$ for each mode order $n$ indicated. The third (fourth) columns show, for each $n$, the stable t-bone (jackknife) mode for $\gamma/g=3.3$ and the corresponding wavefield~\eqref{hSurfer} evaluated at $t=0$. Scale bars denote the capillary wavelength $\lambda_c$. Movies of these t-bone and jackknife modes are shown in Supplemental Video 5.
      }
        \label{JTPlot}
\end{figure*}

In the t-bone [Fig.~\ref{BoundStates}(f)] and jackknife [Fig.~\ref{BoundStates}(g)] modes, two surfers execute circular orbits of different radii around a common center. We locate these modes by solving Eqs.~\eqref{GreenBox2}-\eqref{RedBox} for the three unknowns $d$, $\varphi_1$ and $\varphi_2$. The contour method described in \S\ref{SSec:Prom} is designed for two unknowns and thus cannot be used; we instead use MATLAB's root-finding algorithm to locate some of the modes, and leave the identification of all possible t-bone and jackknife modes for future work. The dependence of $d$, $\varphi_1$ and $\varphi_2$ on the forcing acceleration $\gamma$ is shown in Fig.~\ref{JTPlot}. 
We observe that stable t-bone (jackknife) modes satisfy $\varphi_1\lesssim 0$ ($\varphi_1\lesssim \pi$), and both satisfy $\varphi_2\approx \pi/2$. As with the orbiting modes [Fig.~\ref{OrbPlot}], the distance $d$ between surfers is quantized on the capillary wavelength, with the t-bone modes consistently larger than the jackknife modes. The trajectories and wavefields in the third and fourth columns of Fig.~\ref{JTPlot} are obtained by recasting the solutions in terms of $\bs{x}_1=(\bs{\sigma}-\bs{\delta})/2$ and $\bs{x}_2=(\bs{\sigma}+\bs{\delta})/2$. Specifically, we let $\bs{\sigma}\equiv s( \cos(\omega_0 t+\psi),\sin(\omega_0 t+\psi))$ and find $s$ and $\psi$ by numerically solving the system of equations
\begin{align}
-\tilde{m}s\omega_0^2=\cos(\varphi_2-\psi)+\cos(\varphi_1-\psi),\quad s\omega_0 = \sin(\varphi_2-\psi)+\sin(\varphi_1-\psi),
\end{align}
which are obtained from Eq.~\eqref{CMA}.

\section{Collective modes}\label{Sec:Collective}

\begin{figure*}[ht]
  \centering
    \includegraphics[width=0.65\textwidth]{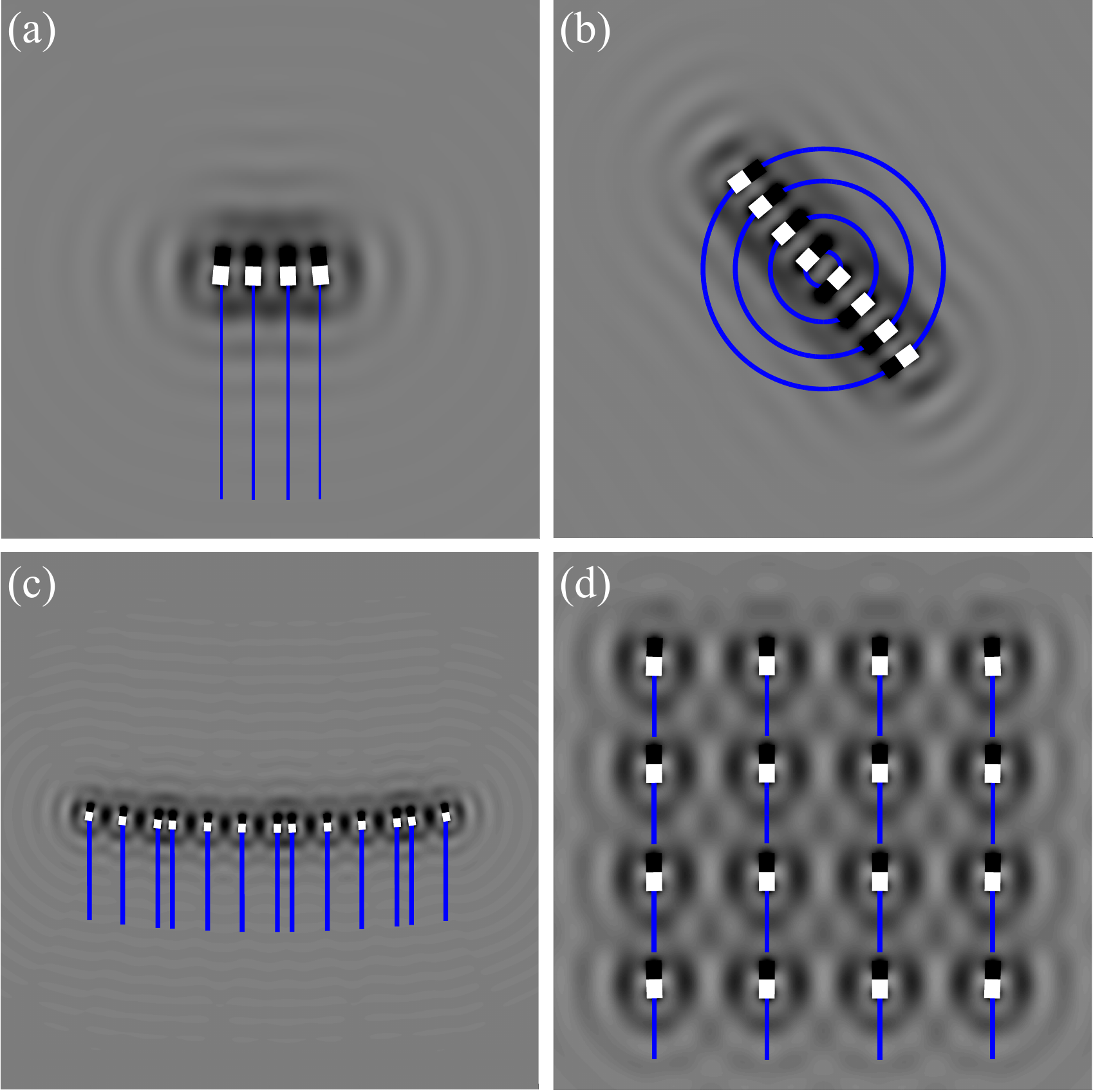}
      \caption{Collective modes of capillary surfers obtained through numerical simulations of Eq.~\eqref{NDimEq}. (a) 4-surfer promenade mode, where the surfers translate at constant velocity and neighbors are separated by approximately one capillary wavelength. (b) 8-surfer super-orbiting mode, where the collective executes uniform circular motion at constant angular frequency and neighbors are separated by approximately one capillary wavelength. (c) Flocking state of thirteen surfers, wherein the collective moves upward with constant velocity. Pairs of surfers are separated by approximately one or two capillary wavelengths. (d) A flocking state of sixteen surfers, in which the collective moves with constant velocity. Neighboring surfers are separated by approximately three capillary wavelengths in both the horizontal and vertical directions. All four modes are obtained for the parameter combination $f=100$ Hz and $\gamma/g=3.3$. These four modes are shown in Supplemental Videos 6 through 9, respectively.
      }
        \label{CollectivePlot}
\end{figure*}

Experiments and simulations of our model~\eqref{NDimParam} show that collections of capillary surfers exhibit novel self-organization phenomena. For example, a many-body promenade mode has been observed in experiment [Fig. 4(a) in~\cite{HoSurfers}] and simulations [Fig.~\ref{CollectivePlot}(a), Supplemental Video 6]. Similarly, simulations are able to reproduce the ``super-orbiting mode" [Fig.~\ref{CollectivePlot}(b), Supplemental Video 7], wherein eight surfers execute orbital motion around a fixed center of mass [Fig. 4(b) in~\cite{HoSurfers}]. Owing to its simplicity, the theoretical model is also able to produce more exotic collective modes that are currently difficult to realize in experiments. For example, Fig.~\ref{CollectivePlot}(c) (Supplemental Video 8) shows an exotic promenade mode of 13 surfers, in which the spacing between neighboring surfers is approximately either one or two capillary wavelengths. This mode may be thus interpreted as an aggregate of $n=1$ and $n=2$ promenade modes [Fig.~\ref{PromPlotsFreq}], and exhibits an example of how the multistable quantized states obtained in \S\ref{Sec:BoundStates} can be used as building blocks for many-body states. Figure~\ref{CollectivePlot}(d) (Supplemental Video 9) shows a similar phenomenon, wherein a square lattice of 16 surfers executes a coherent flocking state with constant velocity. This state may be interpreted as a combination of the $n=3$ tailgating [Fig.~\ref{Fig:ForceCurves}] and promenade modes.

\section{Conclusion}\label{Sec:Conclusion}

We have presented a theoretical model~\eqref{eq:dimmodel} for the dynamics of capillary surfers [Fig.~\ref{Schematic}(a-b)], bodies that self-propel while oscillating at the interface of a fluid bath. The interfacial deformation generated by such a body is calculated by splitting it into static and dynamic contributions, the former resulting from the body’s weight and the latter from the prescribed oscillation of the body at the interface. The static contribution~\eqref{FStatic} to the force is obtained in \S\ref{SSec:Static} by approximating the surfer as a pair of floating discs with unequal masses [Fig.~\ref{Schematic}(c-d)]. The dynamic contribution to the force [Eq.~\eqref{ViscGravForce}, Fig.~\ref{Fig:ViscGrav}(c)] is obtained in \S\ref{SSec:Dynamic} by approximating the surfer as a pair of point sources of weakly viscous gravity-capillary waves, the point-source approximation being required because there does not exist a formula for the dynamic interfacial deformation generated by a finite-sized oscillating body. The resulting formula for the dynamic force is obtained by making use of the results in \S\ref{Sec:Waves}, in which we solved the quasipotential wave model~\eqref{ViscGravPotFlow} and thus derived a formula [Eq.~\eqref{h1AnalGrav}, Fig.~\ref{Fig:ViscGrav}(a-b)] for the small-amplitude (linear) wavefield generated by an oscillating point source. 

The resulting model contains as its only free parameter the speed of a single surfer $U$, which is obtained from experiment~\cite{HoSurfers}. For the case of two surfers, the model recovers the seven bound states observed in experiments~\cite{HoSurfers} [Fig.~\ref{BoundStates}]. We found exact solutions for the head-to-head, back-to-back and tailgating modes in \S\ref{SSec:1DOF} and investigated their stability in Appendix~\ref{App:Stability_Rect}. These solutions are quantized on the capillary wavelength $\lambda_c$, with stable branches of solutions separated by unstable ones [Fig.~\ref{Fig:ForceCurves}]. An exact solution for the promenade mode is found in \S\ref{SSec:Prom}, and the theoretical predictions correctly capture the trends observed in experiment. Moreover, the theoretically predicted dependence of the distance between surfers on the forcing frequency is in excellent quantitative agreement with experiment [Fig.~\ref{PromPlotsFreq}]. However, the predicted dependence of the distance on the forcing acceleration for a fixed value of the forcing frequency ($f=100$ Hz) exhibits small but systematic discrepancies with experiment. This is presumably due to the fact that the quasipotential approximation for the wavefield is valid in the low frequency regime $\epsilon \ll 1$. The model also overpredicts the promenade speed, presumably due to the fact that we neglect modulations in the surfers' vertical dynamics. We also found exact solutions for the orbiting [Fig.~\ref{OrbPlot}], jackknife and t-bone [Fig.~\ref{JTPlot}] modes in \S\ref{SSec:Rot} and investigated their stability in Appendix~\ref{App:Stability_Rot}.

All of the bound states described in \S\ref{Sec:BoundStates} exhibit multistability of a discrete set of interaction states, wherein a number of states quantized on the capillary wavelength may stably coexist for the same experimental parameters. This feature is due to the wave-mediated interactions between surfers, which result in long-range spatially-oscillatory forces defined by alternating regions of attraction and repulsion [Fig.~\ref{Fig:ViscGrav}(c)]. Such interactions give rise to the collective modes shown in Fig.~\ref{CollectivePlot}, which may be viewed as combinations of pairwise bound states.

While the point force approximation is expected to be valid when the distance between surfers is much larger than the surfer's length, many of the bound states and collective modes reported in experiments consist of closely-separated surfers~\cite{HoSurfers}. A promising future direction would thus be to develop a theory for the dynamic deformation generated by a finite-sized body oscillating on a fluid interface. A theory for dynamically floating bodies may also shed light on the propulsion mechanism of surfers, thus allowing us to eliminate the {\it ad hoc} propulsive force $F_p\bs{n}_i$ in our model~\eqref{eq:dimmodel}. Moreover, Fig.~\ref{CollectivePlot} only describes a small sample of the rich variety of collective modes expected to arise in the surfer system. The self-organization and emergent collective behavior exhibited by large populations of surfers will thus be detailed in future work. 



\begin{acknowledgments}
AO acknowledges support from the Simons Foundation (Collaboration Grant for Mathematicians, Award No.
587006) and NSF DMS-2108839. DMH acknowledges support from the Office of Naval Research (ONR N00014-21-1-2816) and the Brown Undergraduate Teaching and Research Award. Special thanks for Prof. Adri Olde Daalhuis for assisting with the argument given in Appendix~\ref{App:FarField}, and to Jack-William Barotta for useful discussions.
\end{acknowledgments}

\appendix
\section{Inviscid linear waves generated by an oscillating point source}\label{App:dCG}

We here derive the linear wave field generated by a point force oscillating harmonically on the free surface of an inviscid fluid bath in the absence of gravity, a problem first considered by De Corato \& Garbin~\cite{de2018capillary}. The derivation proceeds as in \S\ref{Sec:Waves}, with the reciprocal Reynolds number and wave Bond numbers set to zero, $\epsilon=\beta=0$. Equation~\eqref{h1IntGrav} then reads
\begin{align}
h_1(\bs{x})=\frac{F_0}{2\pi\sigma}\int_0^{\infty}\rmd k\,\frac{k^2}{k^3-1}\mathrm{J}_0(kk_cr).
\end{align}
Following Appendix A in~\cite{de2018capillary}, we compute the integral by rewriting the rational function in the integrand above,
\begin{align}
\frac{k^2}{k^3-1}=\frac{1}{3}\left(\frac{1}{k-1}+\frac{1}{k+\varsigma }+\frac{1}{k+\bar{\varsigma}}\right),
\end{align}
and using the fact that~\cite[2.12.3.6, p. 175]{PrudnikovBook}
\begin{align}
\int_0^{\infty}\frac{\mathrm{J}_0(kk_cr)}{k+k_0}\,\rmd k = \frac{\pi}{2}C_0(k_0k_cr)\quad\text{for }k_0\in\mathbb{C} \text{ with }\text{Im}(k_0)\neq 0.\label{BesselID1}
\end{align}
However, the integral $\int_0^{\infty}\mathrm{J}_0(kk_cr)/(k-1)\,\rmd k$ is divergent. To make sense of the integral, we employ the limiting absorption principle and interpret it as the following limit:
\begin{align}
\lim_{\epsilon\rightarrow 0^+}\int_0^{\infty}\frac{\mathrm{J}_0(kk_cr)}{k-1\pm \rmi\epsilon}\,\rmd k &= \lim_{\epsilon\rightarrow 0^+} \frac{\pi}{2}C_0\left(\left(-1\mp \rmi\epsilon\right)k_cr\right)=\lim_{\epsilon\rightarrow 0^+} \frac{\pi}{2}\left[-\mathrm{H}_0\left(\left(1\pm \rmi\epsilon\right)k_cr\right)-\mathrm{Y}_0\left(\left(1\pm \rmi\epsilon\right)k_cr\right)\mp 2\rmi \mathrm{J}_0\left(\left(1\pm \rmi\epsilon\right)k_cr\right)\right]\nonumber \\
&=-\frac{\pi}{2}\left(\mathrm{H}_0(k_cr)+\mathrm{Y}_0(k_cr)\right)\mp \rmi\pi \mathrm{J}_0(k_cr),
\end{align}
where we use the facts~\cite[Eq.~10.11.6]{NIST:DLMF}
\begin{align}
\mathrm{H}_0(-z) = -\mathrm{H}_0(z)\quad\text{and}\quad \mathrm{Y}_0(-z)=\mathrm{Y}_0(z)-2\rmi\,\text{sgn}\left(\text{Im}[z]\right) \mathrm{J}_0(z)\quad\text{for}\quad z\in\mathbb{C}\text{ with }\text{Im}[z]\neq 0.\label{BesselFacts}
\end{align} 
We thus obtain 
\begin{align}
h_1^{\pm}(\bs{x})=\frac{F_0}{12\sigma}\left\{2\,\text{Re}\left[C_0(\varsigma k_cr)\right]-\mathrm{H}_0(k_cr)-\mathrm{Y}_0(k_cr)\mp 2\rmi \mathrm{J}_0(k_cr)\right\}.\label{h1pm}
\end{align}
We note that the imaginary term is missing from Eq. (3.16) in Ref.~\cite{de2018capillary}. 

To choose the correct sign in Eq.~\eqref{h1pm}, we use the Sommerfeld radiation condition, which ensures that the waves propagate outward from the source:
\begin{align}
\lim_{r\rightarrow\infty}\sqrt{r}\left(\pdiff{}{r}+\rmi k_c\right)h_1^{\pm}=0.\label{SommerfeldCond}
\end{align}
Substituting Eq.~\eqref{h1pm} into~\eqref{SommerfeldCond}, we obtain
\begin{align}
\lim_{r\rightarrow\infty}\sqrt{r}\left\{-2\,\text{Re}\left[\varsigma C_1(\varsigma k_cr)\right]+\mathrm{H}_{1}(k_cr)+\mathrm{Y}_1(k_cr)\pm 2\rmi \mathrm{J}_1(k_cr)+\rmi\left[2\,\text{Re}\left[C_0(\varsigma k_cr)\right]-\mathrm{H}_0(k_cr)-\mathrm{Y}_0(k_cr)\mp 2\rmi \mathrm{J}_0(k_cr)\right]\right\}=0,\label{Sommerfeld}
\end{align}
where we use the facts that $\mathrm{Y}_0^{\prime}=-\mathrm{Y}_1$ and $\mathrm{H}_0^{\prime}=\mathrm{H}_{-1}=2/\pi-\mathrm{H}_1$. Using the far field asymptotic results~\cite[Eq.~11.6.1]{NIST:DLMF}
\begin{align}
C_{0}(z)\sim \frac{2}{\pi z}\quad\text{and}\quad C_1(z)\sim \frac{2}{\pi}\left(1+\frac{1}{z^2}\right)\quad\text{as}\quad |z|\rightarrow\infty\quad\text{in}\quad |\arg z|<\pi,\label{C0Asymp}
\end{align}
Eq.~\eqref{Sommerfeld} reduces to
\begin{align}
\lim_{r\rightarrow\infty}\sqrt{r}\left\{\mathrm{Y}_1(k_cr)\pm \rmi \mathrm{J}_1(k_cr)-\rmi\left[\mathrm{Y}_0(k_cr)\pm \rmi \mathrm{J}_0(k_cr)\right]\right\}=0.\label{dCGRad1}
\end{align}
Using the asymptotic forms for the Bessel function,
\begin{align}
\mathrm{J}_{n}(x)\sim\sqrt{\frac{2}{\pi x}}\cos\left(x-\frac{\pi}{4}-\frac{n\pi}{2}\right)\quad \text{and}\quad \mathrm{Y}_{n}(x)\sim\sqrt{\frac{2}{\pi x}}\sin\left(x-\frac{\pi}{4}-\frac{n\pi}{2}\right)\quad\text{as}\quad x\rightarrow\infty,
\end{align}
we deduce that Eq.~\eqref{dCGRad1} is satisfied for the solution with the positive sign; that is, $h_1^+(\bs{x})$ satisfies the radiation condition~\eqref{Sommerfeld}. Using Eq.~\eqref{hFinal}, we conclude that the wavefield has the form
\begin{align}
h(\bs{x},t)=\frac{F_0}{12\sigma}\left\{\left[2\,\text{Re}\left[C_0(\varsigma k_cr)\right]-\mathrm{H}_0(k_cr)-\mathrm{Y}_0(k_cr)\right]\cos\omega t+2\mathrm{J}_0(k_cr)\sin\omega t\right\}.\label{hFinalInviscid}
\end{align}
We note that our result differs from that of Ref.~\cite{de2018capillary} due to the sine-term (see Eq. (3.17) therein). That is, the waveform in Ref.~\cite{de2018capillary} is a standing wave due to the authors' assumption of a reflecting boundary condition at infinity (see Eq. (2.7) therein); however, our radiation condition~\eqref{SommerfeldCond} enforces the requirement that waves propagate outward from the source, which is evident from Supplemental Video 1 (right panel). Moreover, by combining Eqs.~\eqref{h1Approx2} and~\eqref{BesselFacts}, it is evident that the waveform~\eqref{h1AnalGrav} that we derived for weakly viscous gravity-capillary waves reduces to $h_1^+(\bs{x})$ in Eq.~\eqref{h1pm} if the effects of gravity ($\beta=0$) and viscosity ($\epsilon\rightarrow 0$) are neglected.

We conclude by computing the time-averaged force exerted by one oscillating particle on another: specifically, suppose particles (labeled 0 and 1) separated by a distance $r$ exert vertical forces $F_0\cos\omega t$ and $F_1\cos(\omega t+\phi_1)$ on the fluid interface. The force on particle 1 due to the interfacial deformation generated by particle 0 is
\begin{align}
\langle F_1\cos(\omega t+\phi_1)\bs{\nabla}h(\bs{x},t)\rangle=\frac{F_0F_1k_c}{24\sigma}\left\{\left[-2\,\text{Re}\left[\varsigma C_1(\varsigma k_cr)\right]+\mathrm{H}_{-1}(k_cr)+\mathrm{Y}_1(k_cr)\right]\cos\phi_1+2\mathrm{J}_1(k_cr)\sin\phi_1\right\}\hat{\bs{r}},\label{dCGForce}
\end{align}
where $\hat{\bs{r}}$ is a unit vector that points from particle 0 to 1. If the particles oscillate in-phase ($\phi_1=0$) or out-of-phase ($\phi_1=\pi$), we recover the expression derived in Ref.~\cite{de2018capillary} (see Eq. (3.20) therein); however, other phase relationships will result in deviations from that expression owing to the $\mathrm{J}_1$--term in Eq.~\eqref{dCGForce}.

\section{Far-field behavior of the wave field generated by an oscillating point source in the small viscosity limit}\label{App:FarField}

\begin{figure*}[ht]
  \centering
    \includegraphics[width=1\textwidth]{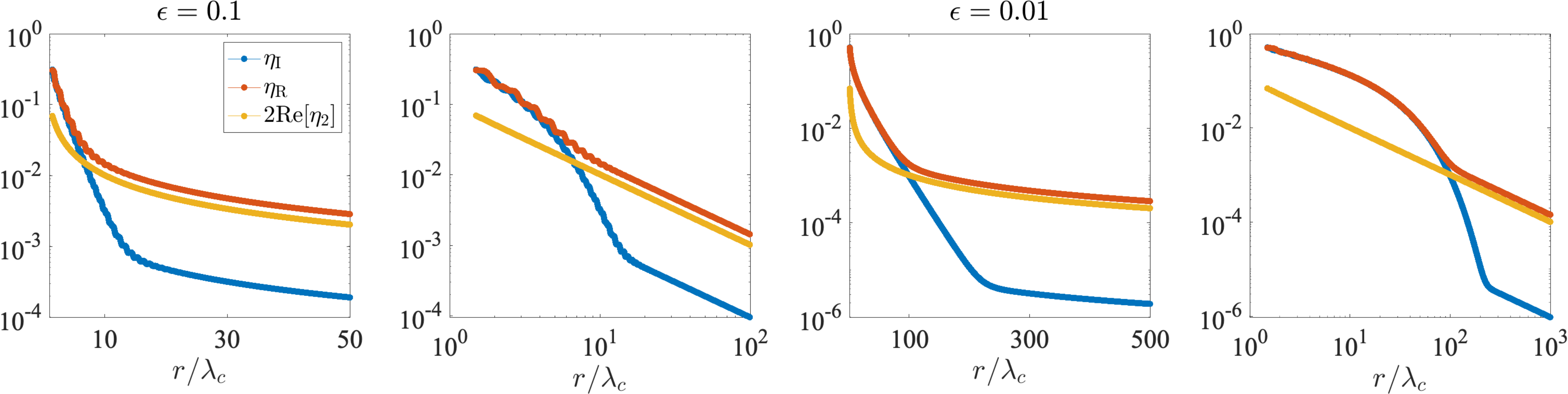}
      \caption{Plots of the functions $\eta_{\text{I}}(r)$, $\eta_{\text{R}}(r)$ and $2\,\text{Re}[\eta_2(r)]$, as defined by Eqs.~\eqref{etaAmps} and~\eqref{h1Approx2}. Panels (a) and (b) correspond to $\epsilon = 0.1$, and (c) and (d) to $\epsilon = 0.01$. In each pair, the panel on the left (right) is on semi-logarithmic (logarithmic) scale to illustrate the far-field behavior of each function.
      }
        \label{EtaFuncs}
\end{figure*}

We now consider the far-field behavior of the wavefield $h_1(r)$ in the regime where viscous effects are small but nonzero ($0 < \epsilon \ll 1$). An approximation of $h_1$ is given by Eq.~\eqref{h1Approx2}, and we wish to compare the magnitudes of the two terms $\eta_1(r)$ and $\eta_2(r)$. We observe that $\text{Re}[\eta_2]$ decreases monotonically in $r$ [Fig.~\ref{EtaFuncs}]. Since the real and imaginary parts of $\eta$ oscillate between positive and negative values, we instead consider their local amplitudes
\begin{align}
\eta_{\text{R}}(r)=\left(\frac{2}{\lambda_c}\int_{r-\lambda_c/2}^{r+\lambda_c/2}\left(\text{Re}[\eta_1(r^{\prime})]\right)^2\,\mathrm{d}r^{\prime}\right)^{1/2}\quad\text{and}\quad\eta_{\text{I}}(r)=\left(\frac{2}{\lambda_c}\int_{r-\lambda_c/2}^{r+\lambda_c/2}\left(\text{Im}[\eta_1(r^{\prime})]\right)^2\,\mathrm{d}r^{\prime}\right)^{1/2},\label{etaAmps}
\end{align}
which are shown in Fig.~\ref{EtaFuncs} for two different values of $\epsilon$. First, we note that $\text{Re}(\eta_2)$ decays algebraically in $r$, as expected from Eq.~\eqref{C0Asymp}, and that $\text{Re}(\eta_2)$ is dominated by $\eta_{\text{R}}$ and $\eta_{\text{I}}$ in the region $1\leq r/\lambda_c\ll 1/\epsilon$. Second, while $\eta_1$ also decays algebraically as $r\rightarrow\infty$, we are interested in its behavior for small $\epsilon$, which corresponds to $\arg z_\epsilon\approx \pi$ for $z_\epsilon = \left(-1+2\rmi\epsilon/3\right)k_cr$. The connection formula~\eqref{BesselFacts} implies that 
 \begin{align}
 \eta_1(r)=-C_0\left[\left(1-\frac{2\rmi\epsilon}{3}\right)k_cr\right]-2\rmi\, \mathrm{H}_0^{(2)}\left[\left(1-\frac{2\rmi\epsilon}{3}\right)k_cr\right],\label{eta1Conn}
 \end{align}
where $\mathrm{H}_0^{(2)}$ is the Hankel function of order zero of the second kind (not to be confused with the Struve function). The Hankel function dominates over $C_0$ for small $\epsilon$ and $k_cr=O(1)$, and its asymptotic behavior is given by~\cite[Eq.~10.17.6]{NIST:DLMF}
\begin{align}
\mathrm{H}_0^{(2)}(z)\sim\sqrt{\frac{2}{\pi z}}\exp\left[-\rmi\left(z-\pi/4\right)\right]\quad\text{as}\quad |z|\rightarrow\infty.\label{HankelAsymp}
\end{align}
From Eqs.~\eqref{eta1Conn} and~\eqref{HankelAsymp}, we conclude that $\eta_1$ decays exponentially in the region $r/\lambda_c=O(1/\epsilon)$ and algebraically thereafter, which is confirmed by Fig.~\ref{EtaFuncs}. We also observe that viscosity damps the waves generated by the point source, since the decay length $3\lambda_c/4\pi\epsilon=3\lambda_c^3\omega/4(2\pi)^3\nu$ is inversely proportional to the viscosity $\nu$.

\section{Linear stability analysis of bound states of surfer pairs}\label{App:Stability}
Here we perform the linear stability analysis of rectilinear (Appendix~\ref{App:Stability_Rect}) and rotating (Appendix~\ref{App:Stability_Rot}) bound states. It is useful to use Eq.~\eqref{SigmaDeltaDef} to write Eq.~\eqref{NDimEqPair} in the form
\begin{subequations}
\label{NDimEqPair2}
\begin{align}
\tilde{m}\ddot{\bs{\sigma}}&=-\dot{\bs{\sigma}}+\bs{n}_1+\bs{n}_2,\label{Pair2A} \\
\tilde{m}\ddot{\bs{\delta}}&=-\dot{\bs{\delta}}+\bs{n}_2-\bs{n}_1-2\tilde{F}_c\left[\mu_+^2\bs{F}_{++}+\mu_{+}\mu_-\left(\bs{F}_{+-}+\bs{F}_{-+}\right)+\mu_-^2\bs{F}_{--}\right],\label{Pair2B} \\
\tilde{m}\tilde{l}\ddot{\theta}_1&=-\tilde{l}\dot{\theta}_1+\tilde{F}_c\bs{n}_1\times\left[\mu_+\left(\bs{F}_{+-}-\bs{F}_{++}\right)-\mu_-\left(\bs{F}_{-+}-\bs{F}_{--}\right)\right],\label{Pair2C}\\
\tilde{m}\tilde{l}\ddot{\theta}_2&=-\tilde{l}\dot{\theta}_2-\tilde{F}_c\bs{n}_2\times\left[\mu_+\left(\bs{F}_{-+}-\bs{F}_{++}\right)-\mu_-\left(\bs{F}_{+-}-\bs{F}_{--}\right)\right]\label{Pair2D},
\end{align}
\end{subequations}
where
\begin{align}
\bs{F}_{pq}=f_{pq}\bs{\delta}_{pq}=\frac{\Phi(|\bs{\delta}_{pq}|)}{|\bs{\delta}_{pq}|}\bs{\delta}_{pq},\quad p,q=+\text{ or }-.
\end{align}
Given a base state $\bs{x}^{\circ}$ and perturbation $\tilde{\bs{x}}$, an object that will show up repeatedly is
\begin{align}
\frac{\Phi(|\bs{x}^\circ+\epsilon\tilde{\bs{x}}|)}{|\bs{x}^\circ+\epsilon\tilde{\bs{x}}|}(\bs{x}^\circ+\epsilon\tilde{\bs{x}})=\frac{\Phi(|\bs{x}^\circ|)}{|\bs{x}^\circ|}\bs{x}^\circ+\epsilon\mathcal{L}(\bs{x}^\circ)\tilde{\bs{x}}+O(\epsilon^2),\quad\text{where}\quad\mathcal{L}(\bs{x}) = 
\frac{\Phi(|\bs{x}|)}{|\bs{x}|}\frac{\bs{x}^{\perp}\bs{x}^{\perp}}{|\bs{x}|^2}+\Phi^{\prime}(|\bs{x}|)\frac{\bs{x}\bs{x}}{|\bs{x}|^2}\label{LId}
\end{align}
and $\bs{x}^{\perp}=(x,y)^{\perp}=(-y,x)$. 
The derivative of the dimensionless force is $\Phi^{\prime}(r) = \alpha f_{\text{s}}^{\prime}(r)+\xi^2f_{\text{d}}^{\prime}(r)$, where, from Eq.~\eqref{FijID_new},
 \begin{align}
f_{\text{s}}^{\prime}(r) = \sqrt{\beta}\mathrm{K}_1^{\prime}(\sqrt{\beta}r)\quad \text{and}\quad f_{\text{d}}^{\prime}(r)=-\sum_{j=1}^4\text{Re}\left[k_j^2\frac{\mathrm{H}_{-1}^{\prime}(-k_jr)+\mathrm{Y}_1^{\prime}(-k_jr)}{1+\beta/3k_j^2+(4/3)\rmi\epsilon/k_j+(4/3)\epsilon^2k_j}\right].\label{PhiPrime}
 \end{align}
To evaluate Eq.~\eqref{PhiPrime}, we use the identities
 \begin{align}
 \mathrm{K}_1^{\prime}(x) = -\frac{1}{2}\left(\mathrm{K}_0(x)+\mathrm{K}_2(x)\right),\quad \mathrm{Y}_1^{\prime}(x) = \frac{1}{2}\left(\mathrm{Y}_0(x)-\mathrm{Y}_2(x)\right),\quad \text{and}\quad \mathrm{H}_{-1}^{\prime}(x) = \frac{1}{\pi x}+\frac{1}{2}\left(\mathrm{H}_{-2}(x)-\mathrm{H}_0(x)\right).
 \end{align}

\subsection{Rectilinear modes}\label{App:Stability_Rect}

We linearize Eq.~\eqref{NDimEqPair2} around the base state $(\bs{x}_i^\circ,\theta_i^\circ)$ for $i = 1,2$, where $\bs{n}_i=(\cos\theta_i,\sin\theta_i)$. To that end, we substitute the expressions $\bs{x}_i=\bs{x}_i^\circ+\epsilon\tilde{\bs{x}}_i$ and $\theta_i = \theta_i^\circ+\epsilon\tilde{\theta}_i$ into Eq.~\eqref{NDimEqPair2} and retain terms at leading order in $\epsilon$. Using the fact that $\bs{n}_i=\bs{n}_i^\circ+\epsilon\bs{n}_i^{\circ\perp}\tilde{\theta}_i+O(\epsilon^2)$, we obtain the linearized equations of motion
\begin{align}
\tilde{m}\dsdiff{\tilde{\bs{\sigma}}}{t}&=-\diff{\tilde{\bs{\sigma}}}{t}+\bs{n}_1^{\circ\perp}\tilde{\theta}_1+\bs{n}_2^{\circ\perp}\tilde{\theta}_2,\nonumber \\
\tilde{m}\dsdiff{\tilde{\bs{\delta}}}{t}&=-\diff{\tilde{\bs{\delta}}}{t}+\bs{n}_2^{\circ\perp}\tilde{\theta}_2-\bs{n}_1^{\circ\perp}\tilde{\theta}_1-2\tilde{F}_c\left[\mu_+^2\mathbb{L}_{++}\tilde{\bs{\delta}}_{++}+\mu_+\mu_-\left(\mathbb{L}_{+-}\tilde{\bs{\delta}}_{+-}+\mathbb{L}_{-+}\tilde{\bs{\delta}}_{-+}\right)+\mu_-^2\mathbb{L}_{--}\tilde{\bs{\delta}}_{--}\right],\nonumber \\
\tilde{m}\tilde{l}\,\dsdiff{\tilde{\theta}_1}{t}&=-\tilde{l}\diff{\tilde{\theta}_1}{t}+\tilde{F}_c\left\{\bs{n}_1^{\circ}\times\left[\mu_+\left(\mathbb{L}_{+-}\tilde{\bs{\delta}}_{+-}-\mathbb{L}_{++}\tilde{\bs{\delta}}_{++}\right)-\mu_-\left(\mathbb{L}_{-+}\tilde{\bs{\delta}}_{-+}-\mathbb{L}_{--}\tilde{\bs{\delta}}_{--}\right)\right]\right.\nonumber \\
&\phantom{=}\left.+\tilde{\theta}_1\bs{n}_1^{\circ\perp}\times\left[\mu_+\left(\bs{F}_{+-}^{\circ}-\bs{F}_{++}^{\circ}\right)-\mu_-\left(\bs{F}_{-+}^{\circ}-\bs{F}_{--}^{\circ}\right)\right]\right\},\nonumber \\
\tilde{m}\tilde{l}\,\dsdiff{\tilde{\theta}_2}{t}&=-\tilde{l}\diff{\tilde{\theta}_2}{t}-\tilde{F}_c\left\{\bs{n}_2^{\circ}\times\left[\mu_+\left(\mathbb{L}_{-+}\tilde{\bs{\delta}}_{-+}-\mathbb{L}_{++}\tilde{\bs{\delta}}_{++}\right)-\mu_-\left(\mathbb{L}_{+-}\tilde{\bs{\delta}}_{+-}-\mathbb{L}_{--}\tilde{\bs{\delta}}_{--}\right)\right]\right.\nonumber \\
&\phantom{=}\left.+\tilde{\theta}_2\bs{n}_2^{\circ\perp}\times\left[\mu_+\left(\bs{F}_{-+}^{\circ}-\bs{F}_{++}^{\circ}\right)-\mu_-\left(\bs{F}_{+-}^{\circ}-\bs{F}_{--}^{\circ}\right)\right]\right\}.
\label{LinStabSurf}
\end{align}
Here, $\bs{\delta}_{pq}=\bs{\delta}_{pq}^{\circ}+\epsilon\tilde{\bs{\delta}}_{pq}+O(\epsilon^2)$, $\bs{F}_{pq}^\circ =  \Phi(|\bs{\delta}_{pq}^\circ|)\bs{\delta}_{pq}^\circ/|\bs{\delta}_{pq}^\circ|$ and $\mathbb{L}_{pq}=\mathcal{L}(\bs{\delta}_{pq}^\circ)$, where, from Eq.~\eqref{deltapm},
\begin{align}
\tilde{\bs{\delta}}_{++}&=\tilde{\bs{\delta}}-\mu_-\tilde{l}\left(\bs{n}_2^{\circ\perp}\tilde{\theta}_2-\bs{n}_1^{\circ\perp}\tilde{\theta}_1\right),\quad \tilde{\bs{\delta}}_{--}=\tilde{\bs{\delta}}+\mu_+\tilde{l}\left(\bs{n}_2^{\circ\perp}\tilde{\theta}_2-\bs{n}_1^{\circ\perp}\tilde{\theta}_1\right),\nonumber \\
\tilde{\bs{\delta}}_{+-}&=\tilde{\bs{\delta}}-\tilde{l}\left(\mu_-\bs{n}_2^{\circ\perp}\tilde{\theta}_2+\mu_+\bs{n}_1^{\circ\perp}\tilde{\theta}_1\right),\quad \tilde{\bs{\delta}}_{-+}=\tilde{\bs{\delta}}+\tilde{l}\left(\mu_+\bs{n}_2^{\circ\perp}\tilde{\theta}_2+\mu_-\bs{n}_1^{\circ\perp}\tilde{\theta}_1\right).
\end{align}
We note that Eq.~\eqref{LinStabSurf} 
is independent of the (rescaled) center of mass $\tilde{\bs{\sigma}}$ due to translation invariance of the governing equations. Equation~\eqref{LinStabSurf} 
may thus be written in the matrix form (dropping the tildes)
\begin{align}
\diff{\bs{z}}{t}=\mathcal{M}\bs{z},\quad\text{where}\quad \bs{z}=\begin{pmatrix} \bs{\delta} & \dot{\bs{\sigma}} & \dot{\bs{\delta}} & \theta_1 & \theta_2 & \omega_1 & \omega_2\end{pmatrix}^{\text{T}}\quad\text{and}\quad\mathcal{M}=\begin{pmatrix} \mathbb{Z} & \mathbb{Z} & \mathbb{I} & \mathbb{Z} &  \mathbb{Z} \\  \mathbb{Z} & -\mathbb{I}/\tilde{m} & \mathbb{Z} & \mathbb{N} & \mathbb{Z} \\ \mathbb{F}_1 & \mathbb{Z} & -\mathbb{I}/\tilde{m} & \mathbb{F}_2 & \mathbb{Z} \\    \mathbb{Z} &  \mathbb{Z} &  \mathbb{Z} &  \mathbb{Z} &  \mathbb{I} \\    \mathbb{T}_1 &  \mathbb{Z} &  \mathbb{Z} &  \mathbb{T}_2 & - \mathbb{I}/\tilde{m} \end{pmatrix}.\label{RectLinEqs}
\end{align}
Here, $\mathbb{Z}$ and $\mathbb{I}$ are the $2\times 2$ zero and identity matrices, respectively. The $2\times 2$ matrix $\mathbb{F}_1$ is defined as
\begin{align}
\mathbb{F}_1&= -\frac{2\tilde{F}_c}{\tilde{m}}\left[\mu_+^2\mathbb{L}_{++}+\mu_+\mu_-\left(\mathbb{L}_{+-}+\mathbb{L}_{-+}\right)+\mu_-^2\mathbb{L}_{--}\right],
\end{align}
and the $2\times 2$ matrices
\begin{align}
\mathbb{N}&=\frac{1}{\tilde{m}}\begin{pmatrix} \bs{n}_1^{\circ\perp} & \bs{n}_2^{\circ\perp}\end{pmatrix},\quad  \mathbb{F}_2=\begin{pmatrix} \bs{m}_1 & \bs{m}_2\end{pmatrix},\quad\mathbb{T}_1=\begin{pmatrix} \bs{m}_3^{\text{T}} \\ \bs{m}_4^{\text{T}}\end{pmatrix}\quad\text{and}\quad
 \mathbb{T}_2=\begin{pmatrix} m_5 & m_6 \\ m_6 & m_7\end{pmatrix}
\end{align}
are comprised of the elements
\begin{align}
\bs{m}_1&=-\left(\mathbb{I}+2\tilde{F}_c\tilde{l}\mu_+\mu_-\left[\mu_+(\mathbb{L}_{++}-\mathbb{L}_{+-})+\mu_-(\mathbb{L}_{-+}-\mathbb{L}_{--})\right]\right)\frac{\bs{n}_1^{\circ\perp}}{\tilde{m}}
,\nonumber \\
\bs{m}_2&=\left(\mathbb{I}+2\tilde{F}_c\tilde{l}\mu_+\mu_-\left[\mu_+(\mathbb{L}_{++}-\mathbb{L}_{-+})+\mu_-(\mathbb{L}_{+-}-\mathbb{L}_{--})\right]\right)\frac{\bs{n}_2^{\circ\perp}}{\tilde{m}}
,\nonumber \\
\bs{m}_3^{\text{T}}&=\frac{\tilde{F}_c}{\tilde{m}\tilde{l}}\left(\bs{n}_1^{\circ\perp}\right)^{\text{T}}\left[\mu_+\left(\mathbb{L}_{+-}-\mathbb{L}_{++}\right)-\mu_-\left(\mathbb{L}_{-+}-\mathbb{L}_{--}\right)\right],
\nonumber \\
\bs{m}_4^{\text{T}}&=-\frac{\tilde{F}_c}{\tilde{m}\tilde{l}}\left(\bs{n}_2^{\circ\perp}\right)^\text{T}\left[\mu_+\left(\mathbb{L}_{-+}-\mathbb{L}_{++}\right)-\mu_-\left(\mathbb{L}_{+-}-\mathbb{L}_{--}\right)\right],
\nonumber \\
m_5&=\frac{\tilde{F}_c}{\tilde{m}\tilde{l}}\left\{\tilde{l}\bs{n}_1^{\circ}\times\left[-\mu_+\left(\mu_+\mathbb{L}_{+-}+\mu_-\mathbb{L}_{++}\right)-\mu_-\left(\mu_-\mathbb{L}_{-+}+\mu_+\mathbb{L}_{--}\right)\right]\bs{n}_1^{\circ\perp}\right.\nonumber \\
&\phantom{=}\left.+\bs{n}_1^{\circ\perp}\times\left[\mu_+\left(\bs{F}_{+-}^{\circ}-\bs{F}_{++}^{\circ}\right)-\mu_-\left(\bs{F}_{-+}^{\circ}-\bs{F}_{--}^{\circ}\right)\right]\right\}
,\nonumber \\
m_6&=\frac{\tilde{F}_c}{\tilde{m}}\mu_+\mu_-\bs{n}_1^{\circ}\times\left(-\mathbb{L}_{+-}+\mathbb{L}_{++}-\mathbb{L}_{-+}+\mathbb{L}_{--}\right)\bs{n}_2^{\circ\perp}
,\nonumber \\
m_7&=-\frac{\tilde{F}_c}{\tilde{m}\tilde{l}}\left\{\tilde{l}\bs{n}_2^{\circ}\times\left[\mu_+\left(\mu_+\mathbb{L}_{-+}+\mu_-\mathbb{L}_{++}\right)+\mu_-\left(\mu_-\mathbb{L}_{+-}+\mu_+\mathbb{L}_{--}\right)\right]\bs{n}_2^{\circ\perp}\right.\nonumber \\
&\phantom{=}\left.+\bs{n}_2^{\circ\perp}\times\left[\mu_+\left(\bs{F}_{-+}^{\circ}-\bs{F}_{++}^{\circ}\right)-\mu_-\left(\bs{F}_{+-}^{\circ}-\bs{F}_{--}^{\circ}\right)\right]\right\}.
\end{align}
For each of the rectilinear bound states considered in this paper, the matrix $\mathcal{M}$ has a zero eigenvalue due to the solution's rotational invariance (\S\ref{SSSec:RotInvRect}). The stability of the bound state is thus determined by the remaining eigenvalues; a solution is stable if all of the eigenvalues have negative real part, and is unstable otherwise.

\subsubsection{Rotational invariance of rectilinear bound states}\label{SSSec:RotInvRect}

We proceed by showing that the vector
\begin{align}
\bs{v}=\begin{pmatrix} \bs{\delta}^{\circ\perp} &  \bs{n}_1^{\circ\perp}+\bs{n}_2^{\circ\perp} & \bs{0} &\bs{1} & \bs{0}\end{pmatrix}^{\text{T}},\quad\text{where}\quad\bs{0}=(0,0)\text{ and }\bs{1}=(1,1)
\end{align}
is in the nullspace of $\mathcal{M}$, due to the invariance of the governing equations under rotation. The vectors $\bs{v}$ and $\mathcal{M}\bs{v}$ are in $\mathbb{R}^{10}$ and may be viewed as lists with five entries in $\mathbb{R}^2$. We have
\begin{align}
\mathcal{M}\bs{v}=\begin{pmatrix} \bs{0} & \bs{0} & \mathbb{F}_1\bs{\delta}^{\circ\perp}+\mathbb{F}_2\bs{1} & \bs{0} & \mathbb{T}_1\bs{\delta}^{\circ\perp}+\mathbb{T}_2\bs{1} \end{pmatrix}^{\text{T}},\label{MvEq}
\end{align}
where the second entry vanishes because $\mathbb{N}\bs{1}= \left(\bs{n}_1^{\circ\perp}+\bs{n}_2^{\circ\perp} \right)/\tilde{m}$. The third entry is, after some algebra and using the fact that $\mathbb{L}_{pq}\bs{\delta}_{pq}^{\circ\perp}=\bs{F}_{pq}^{\circ\perp}$,
\begin{align}
\mathbb{F}_1\bs{\delta}^{\circ\perp}+\bs{m}_1+\bs{m}_2&=\frac{\bs{n}_2^{\circ\perp}-\bs{n}_1^{\circ\perp}}{\tilde{m}}-\frac{2\tilde{F}_c}{\tilde{m}}\left[\mu_+^2\mathbb{L}_{++}\bs{\delta}_{++}^{\circ\perp}+\mu_+\mu_-\left(\mathbb{L}_{+-}\bs{\delta}_{+-}^{\circ\perp}+\mathbb{L}_{-+}\bs{\delta}_{-+}^{\circ\perp}\right)+\mu_-^2\mathbb{L}_{--}\bs{\delta}_{--}^{\circ\perp}\right]\nonumber \\
&=\frac{\bs{n}_2^{\circ\perp}-\bs{n}_1^{\circ\perp}}{\tilde{m}}-\frac{2\tilde{F}_c}{\tilde{m}}\left[\mu_+^2\bs{F}_{++}^{\circ\perp}+\mu_+\mu_-\left(\bs{F}_{+-}^{\circ\perp}+\bs{F}_{-+}^{\circ\perp}\right)+\mu_-^2\bs{F}_{--}^{\circ\perp}\right],\label{InvEntry3}
\end{align}
which is zero by Eq.~\eqref{Pair2B}. The last entry in Eq.~\eqref{MvEq} is
\begin{align}
\mathbb{T}_1\bs{\delta}^{\circ\perp}+\mathbb{T}_2\bs{1}=\begin{pmatrix}\bs{m}_3\cdot\bs{\delta}^{\circ\perp}+m_5+m_6 \\ \bs{m}_4\cdot\bs{\delta}^{\circ\perp}+m_6+m_7\end{pmatrix},
\end{align}
where
\begin{align}
\bs{m}_3\cdot\bs{\delta}^{\circ\perp}+m_5+m_6 &=\frac{\tilde{F}_c}{\tilde{m}\tilde{l}}\left\{\bs{n}_1^\circ\times\left(-\mu_+\mathbb{L}_{++}\bs{\delta}_{++}^{\circ\perp}+\mu_-\mathbb{L}_{--}\bs{\delta}_{--}^{\circ\perp}+\mu_+\mathbb{L}_{+-}\bs{\delta}_{+-}^{\circ\perp}-\mu_-\mathbb{L}_{-+}\bs{\delta}_{-+}^{\circ\perp}\right)\right.\nonumber \\
&\phantom{=}+\left.\bs{n}_1^{\circ\perp}\times\left[\mu_+\left(\bs{F}_{+-}^{\circ}-\bs{F}_{++}^{\circ}\right)-\mu_-\left(\bs{F}_{-+}^{\circ}-\bs{F}_{--}^{\circ}\right)\right]\right\}=0.
\end{align}
A similar argument shows that $\bs{m}_4\cdot\bs{\delta}^{\circ\perp}+m_6+m_7 =0$, which completes the proof.
\subsection{Rotating modes}\label{App:Stability_Rot}

To assess the stability of rotating bound states, we use the results from \S\ref{App:Stability_Rect}. Substituting the rotating base state solutions directly into Eq.~\eqref{RectLinEqs} would result in a system of equations with time-varying coefficients, so we first transform Eq.~\eqref{RectLinEqs} into a frame rotating with the orbital frequency $\omega_0$. To that end, we let $\hat{\bs{r}}_0=(\cos\omega_0 t,\sin\omega_0 t)$ and $\hat{\bs{\theta}}_0=(-\sin\omega_0 t,\cos\omega_0 t)$, and define the $2\times 2$ matrix $\bs{\Omega}=\begin{pmatrix}\hat{\bs{r}}_0 &\hat{\bs{\theta}}_0\end{pmatrix}$. Since $\diff{}{t}(\bs{\Omega}\bs{v})=\dot{\bs{\Omega}}\bs{v}+\bs{\Omega}\dot{\bs{v}}$, we define the transformed vector $\tilde{z}$ by $\mathcal{R}\tilde{\bs{z}}=\bs{z}$, where $\mathcal{R}$ is the $10\times 10$ matrix
\begin{align}
\mathcal{R}=\begin{pmatrix} \bs{\Omega} & \mathbb{Z} & \mathbb{Z} & \mathbb{Z} & \mathbb{Z}  \\ \mathbb{Z} & \bs{\Omega} & \mathbb{Z} & \mathbb{Z} & \mathbb{Z}  \\ \dot{\bs{\Omega}}  & \mathbb{Z} & \bs{\Omega} & \mathbb{Z} & \mathbb{Z} \\ \mathbb{Z} & \mathbb{Z} & \mathbb{Z} & \mathbb{I} & \mathbb{Z}  \\   \mathbb{Z} & \mathbb{Z} & \mathbb{Z} & \mathbb{Z} & \mathbb{I}\end{pmatrix}.
\end{align}
The linearized equations $\diff{}{t}\bs{z}=\mathcal{M}\bs{z}$ transform into $\diff{}{t}\tilde{\bs{z}}=\widetilde{\mathcal{M}}\tilde{\bs{z}}$, where
\begin{align}
\widetilde{\mathcal{M}}=\mathcal{R}^{-1}\left(\mathcal{M}\mathcal{R}-\dot{\mathcal{R}}\right)=\begin{pmatrix} \mathbb{Z} & \mathbb{Z} & \mathbb{I} & \mathbb{Z} & \mathbb{Z} \\ \mathbb{Z} & -\mathbb{I}/\tilde{m}+\omega_0\mathbb{J} & \mathbb{Z} & \bs{\Omega}^{\text{T}}\mathbb{N} & \mathbb{Z} \\ \omega_0^2\mathbb{I}+\bs{\Omega}^{\text{T}}\mathbb{F}_1\bs{\Omega}+\omega_0\mathbb{J}/\tilde{m} & \mathbb{Z} & 2\omega_0\mathbb{J}-\mathbb{I}/\tilde{m} & \bs{\Omega}^{\text{T}}\mathbb{F}_2 & \mathbb{Z} \\ \mathbb{Z} & \mathbb{Z} & \mathbb{Z} & \mathbb{Z} & \mathbb{I} \\ \mathbb{T}_1\bs{\Omega} & \mathbb{Z} & \mathbb{Z} & \mathbb{T}_2 & -\mathbb{I}/\tilde{m}\end{pmatrix} 
\end{align}
and $\mathbb{J}=\begin{pmatrix} 0 & 1 \\ -1 & 0\end{pmatrix}$. As with the rectilinear modes, the matrix $\widetilde{\mathcal{M}}$ has a zero eigenvalue due to the invariance of the orbital solutions under rotation (\S\ref{SSSec:RotInvRot}). The stability of the solutions is thus determined by the remaining eigenvalues; a solution is stable if all of the eigenvalues have negative real part, and is unstable otherwise.

\subsubsection{Rotational invariance of rotating modes}\label{SSSec:RotInvRot}

Using arguments analogous to those in \S\ref{SSSec:RotInvRect}, we show that
\begin{align}
\bs{v}=\begin{pmatrix} \bs{\Omega}^{\text{T}}\bs{\delta}^{\circ\perp} & -\left(\omega_0\mathbb{J}-\frac{1}{\tilde{m}}\mathbb{I}\right)^{-1}\bs{\Omega}^{\text{T}}\mathbb{N}\bs{1} & \bs{0} & \bs{1} & \bs{0}\end{pmatrix}^{\text{T}}\label{OrbNullVec}
\end{align}
is in the nullspace of the matrix $\widetilde{\mathcal{M}}$. The first, second and fourth entries of the product $\widetilde{M}\bs{v}$ are identically zero. The third entry is
\begin{align}
\left(\omega_0^2\mathbb{I}+\bs{\Omega}^{\text{T}}\mathbb{F}_1\bs{\Omega}+\frac{\omega_0\mathbb{J}}{\tilde{m}}\right)\bs{\Omega}^{\text{T}}\bs{\delta}^{\circ\perp}+\bs{\Omega}^{\text{T}}\mathbb{F}_2\bs{1}&=\bs{\Omega}^{\text{T}}\left[\left(\omega_0^2\mathbb{I}+\frac{\omega_0\mathbb{J}}{\tilde{m}}+\mathbb{F}_1\right)\bs{\delta}^{\circ\perp}+\mathbb{F}_2\bs{1}\right]\nonumber \\
&=\bs{\Omega}^{\text{T}}\left[\left(\omega_0^2\mathbb{I}+\frac{\omega_0\mathbb{J}}{\tilde{m}}+\mathbb{F}_1\right)\bs{\delta}^{\circ\perp}+\bs{m}_1+\bs{m}_2\right].
\end{align}
Using Eq.~\eqref{InvEntry3}, the term in the square brackets may be simplified to
\begin{align}
\omega_0^2\bs{\delta}^{\circ\perp}+\frac{\omega_0}{\tilde{m}}\bs{\delta}^{\circ}+\frac{\bs{n}_2^{\circ\perp}-\bs{n}_1^{\circ\perp}}{\tilde{m}}-\frac{2\tilde{F}_c}{\tilde{m}}\left[\mu_+^2\bs{F}_{++}^{\circ\perp}+\mu_+\mu_-\left(\bs{F}_{+-}^{\circ\perp}+\bs{F}_{-+}^{\circ\perp}\right)+\mu_-^2\bs{F}_{--}^{\circ\perp}\right],
\end{align}
which is zero by Eq.~\eqref{Pair2B}. The fifth entry in $\widetilde{M}\bs{v}$ is $\mathbb{T}_1\bs{\delta}^{\circ\perp}+\mathbb{T}_2\bs{1}$, which is zero as shown in \S\ref{SSSec:RotInvRect}.

\begin{figure*}[ht]
  \centering
    \includegraphics[width=0.5\textwidth]{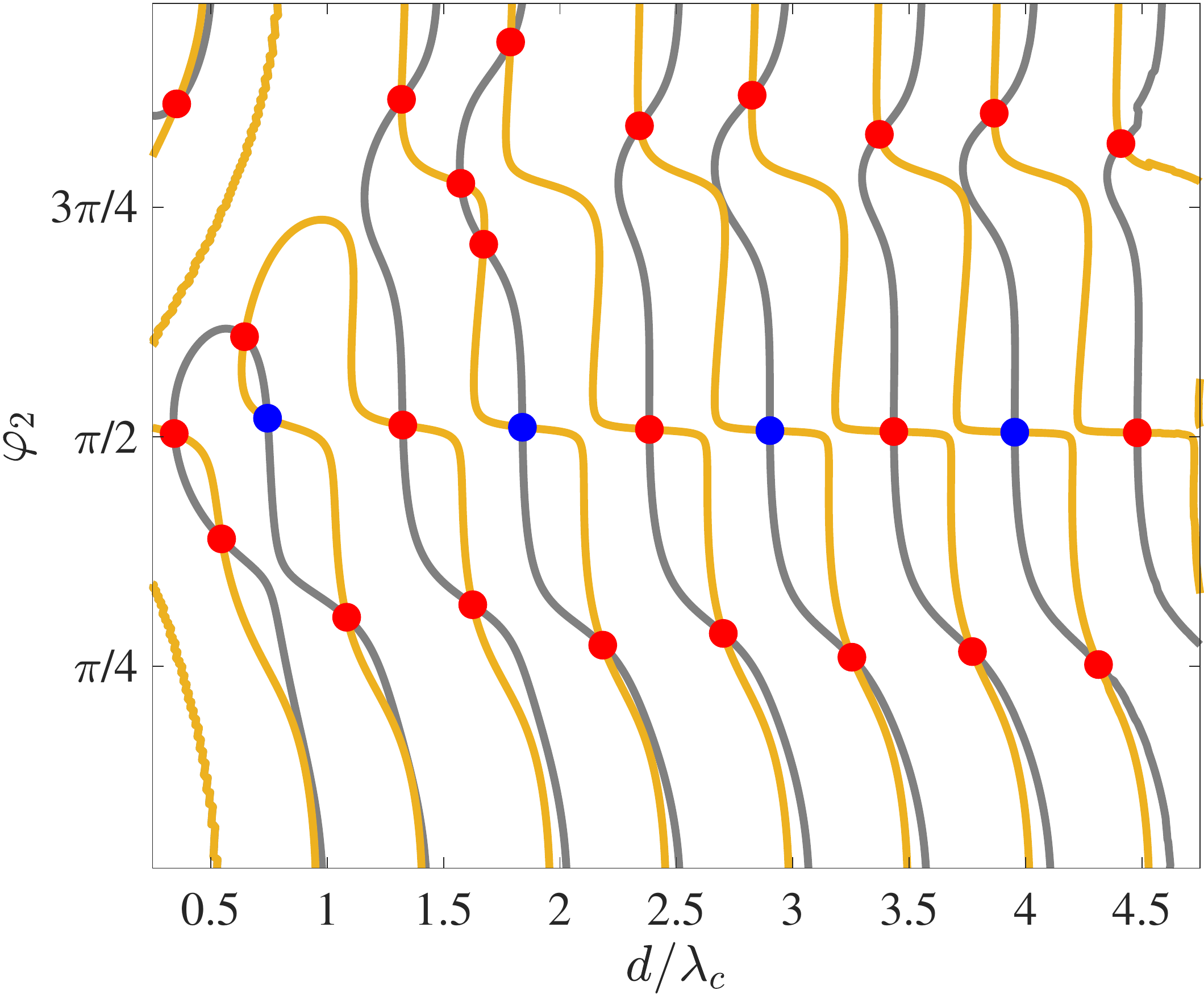}
      \caption{Illustration of the numerical method used to locate promenade mode solutions, as described in \S\ref{SSec:Prom} of the Main Text. The curves are zero-contours of the functions $F_{\text{P}}(d,\varphi_2)$ 
       (gray) and $T_{\text{P}}(d,\varphi_2)$ 
       (yellow), as defined in Eq.~\eqref{PromEqs}, where $d$ is the distance between surfers and $\varphi_2$ 
       the orientation angle. Equilibrium solutions are given by the intersections of the contours, with stable (unstable) points marked in blue (red). The parameters are given in Table~\ref{tab:param}, with $\gamma=3.3g$ and $f = 100$ Hz.   }
        \label{PromContour}
\end{figure*}

\newpage 
\section*{Supplementary Videos}

\begin{figure*}[ht]
  \centering
    \includegraphics[width=0.5\textwidth]{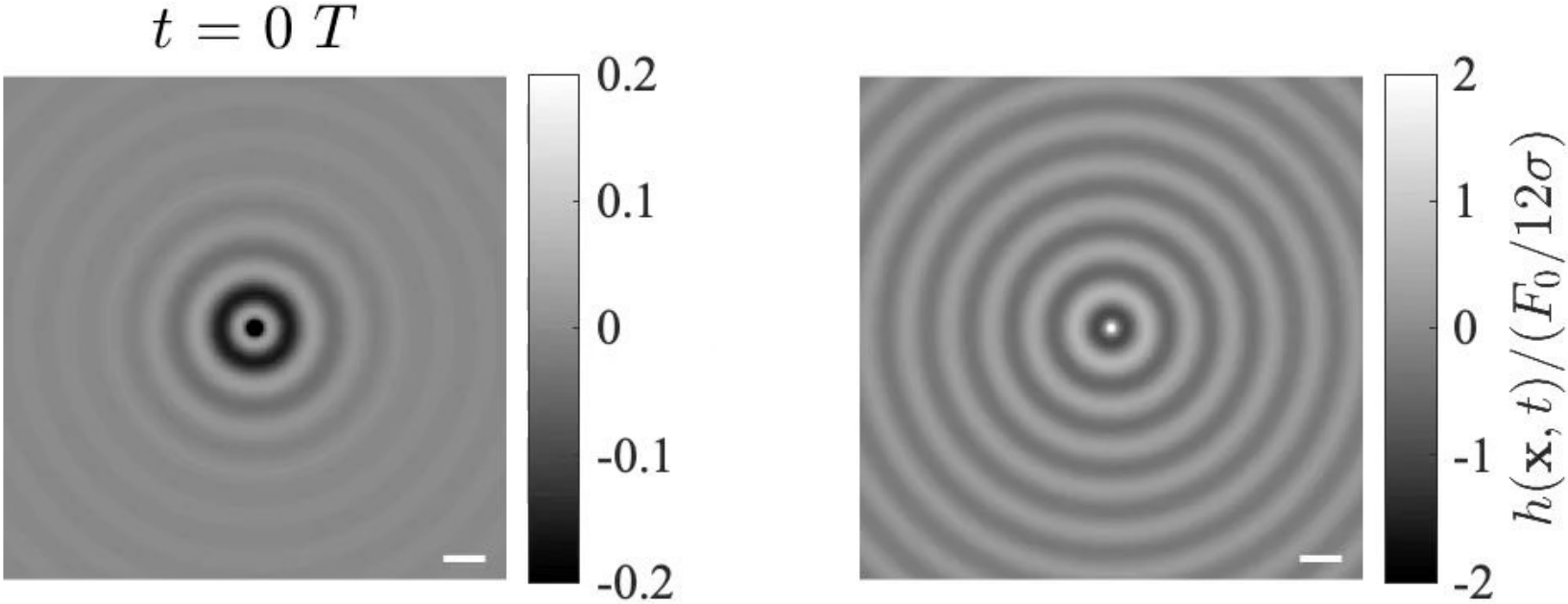} \\ [1mm]
  \raggedright { {\bf Video 1:} Left panel shows the weakly viscous gravity-capillary wavefield~\eqref{hFinal} corresponding to a point particle oscillating at the origin with period $T = 1/f$. Right panel shows the inviscid capillary wavefield derived in Eq.~\eqref{hFinalInviscid}. The scale bars denote the capillary wavelength $\lambda_c$. The parameters correspond to those given in Table~\ref{tab:param}, with forcing frequency $f = 100$ Hz.}
\end{figure*}

\begin{figure*}[ht]
  \centering
    \includegraphics[width=0.5\textwidth]{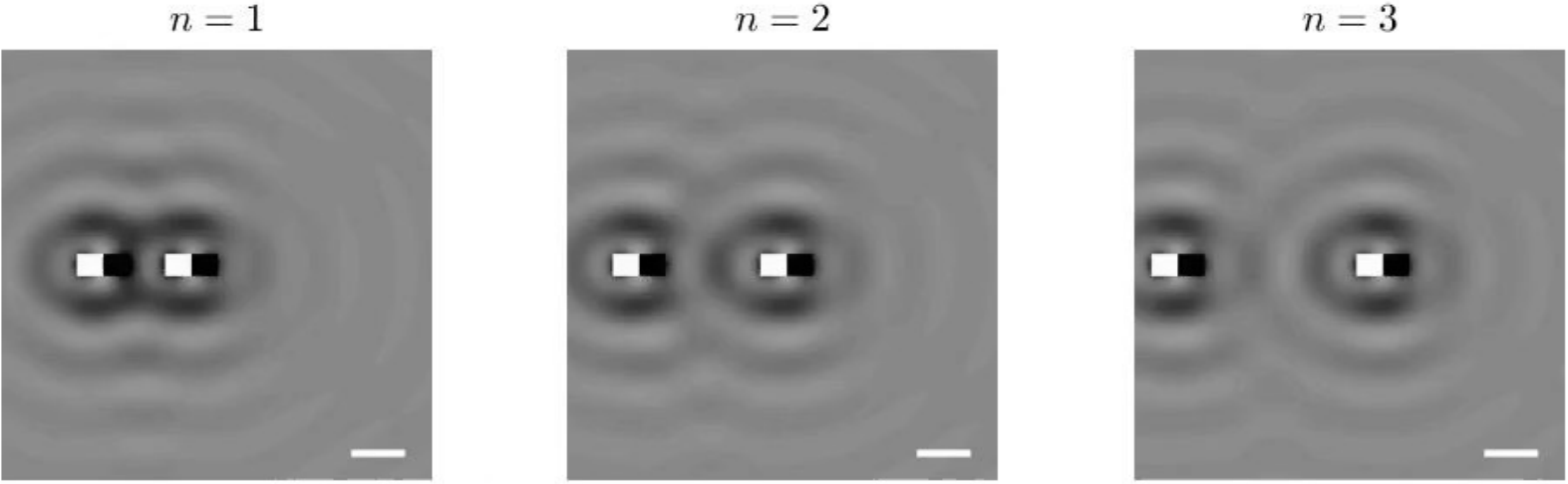} \\ [1mm]
  \raggedright { {\bf Video 2:} First three tailgating modes and corresponding wavefields for $f = 100$ Hz and $\gamma/g = 3.3$, as shown in Fig.~\ref{Fig:ForceCurves}. The scale bars denote the capillary wavelength $\lambda_c$.}
\end{figure*}

\begin{figure*}[ht]
  \centering
    \includegraphics[width=0.5\textwidth]{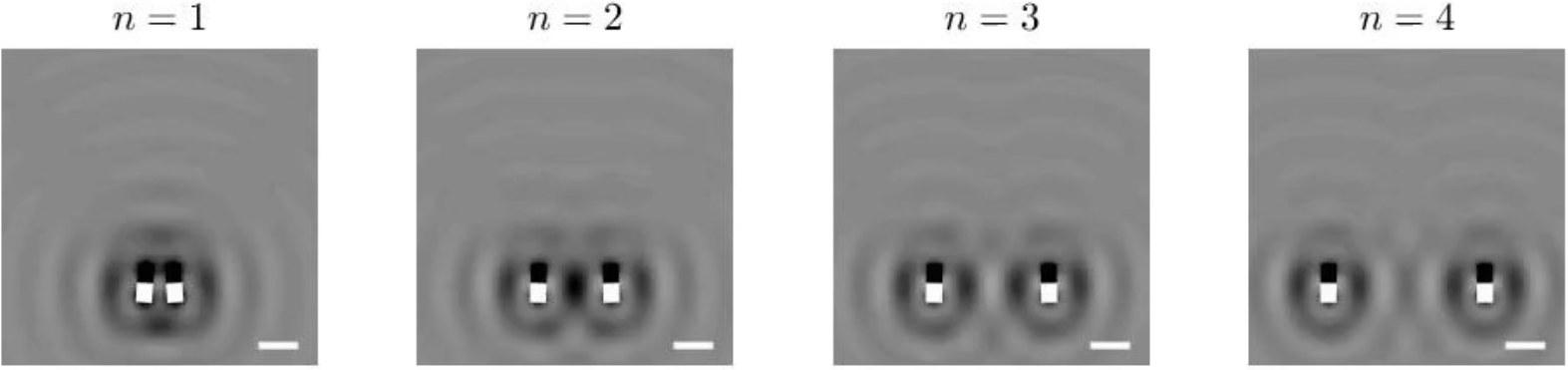} \\ [1mm]
  \raggedright { {\bf Video 3:} First four promenade modes and corresponding wavefields for $f = 100$ Hz and $\gamma/g = 3.3$, as shown in Fig.~\ref{PromPlotsFreq}. The scale bars denote the capillary wavelength $\lambda_c$.}
\end{figure*}

\begin{figure*}[ht]
  \centering
    \includegraphics[width=0.5\textwidth]{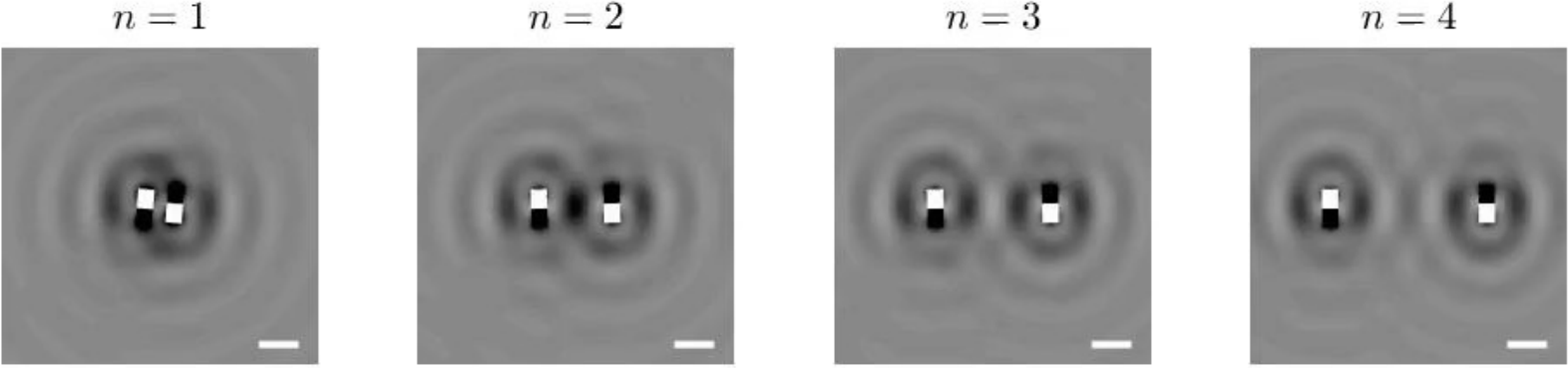} \\ [1mm]
  \raggedright { {\bf Video 4:} First four orbiting modes and corresponding wavefields for $f = 100$ Hz and $\gamma/g = 3.3$, as shown in Fig.~\ref{OrbPlot}. The scale bars denote the capillary wavelength $\lambda_c$.}
\end{figure*}

\begin{figure*}[ht]
  \centering
    \includegraphics[width=0.5\textwidth]{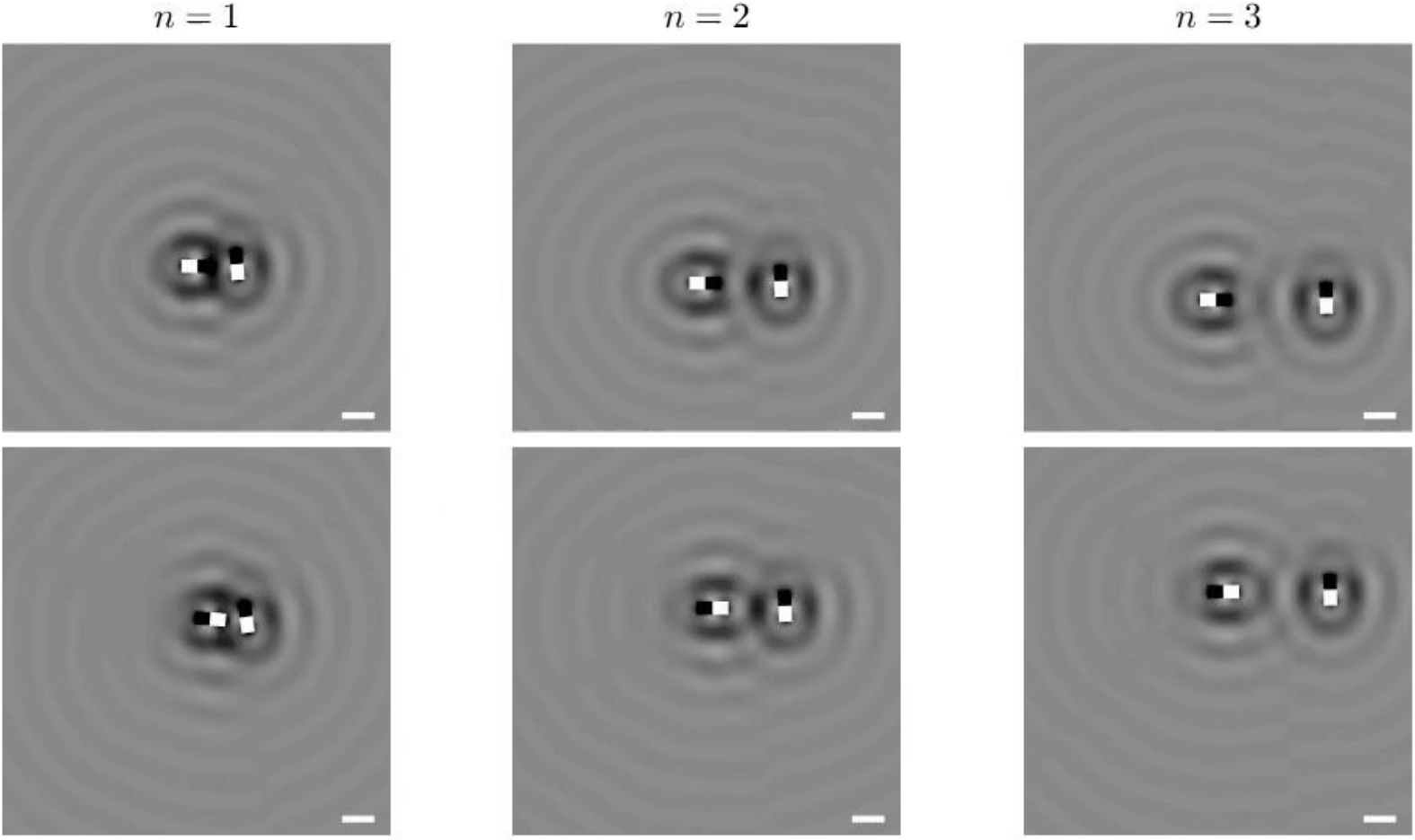} \\ [1mm]
  \raggedright { {\bf Video 5:} Top (bottom) row shows the first three t-bone (jackknife) modes and corresponding wavefields for $f = 100$ Hz and $\gamma/g = 3.3$, as shown in Fig.~\ref{JTPlot}. The scale bars denote the capillary wavelength $\lambda_c$.}
\end{figure*}

\begin{figure*}[ht]
  \centering
    \includegraphics[width=0.2\textwidth]{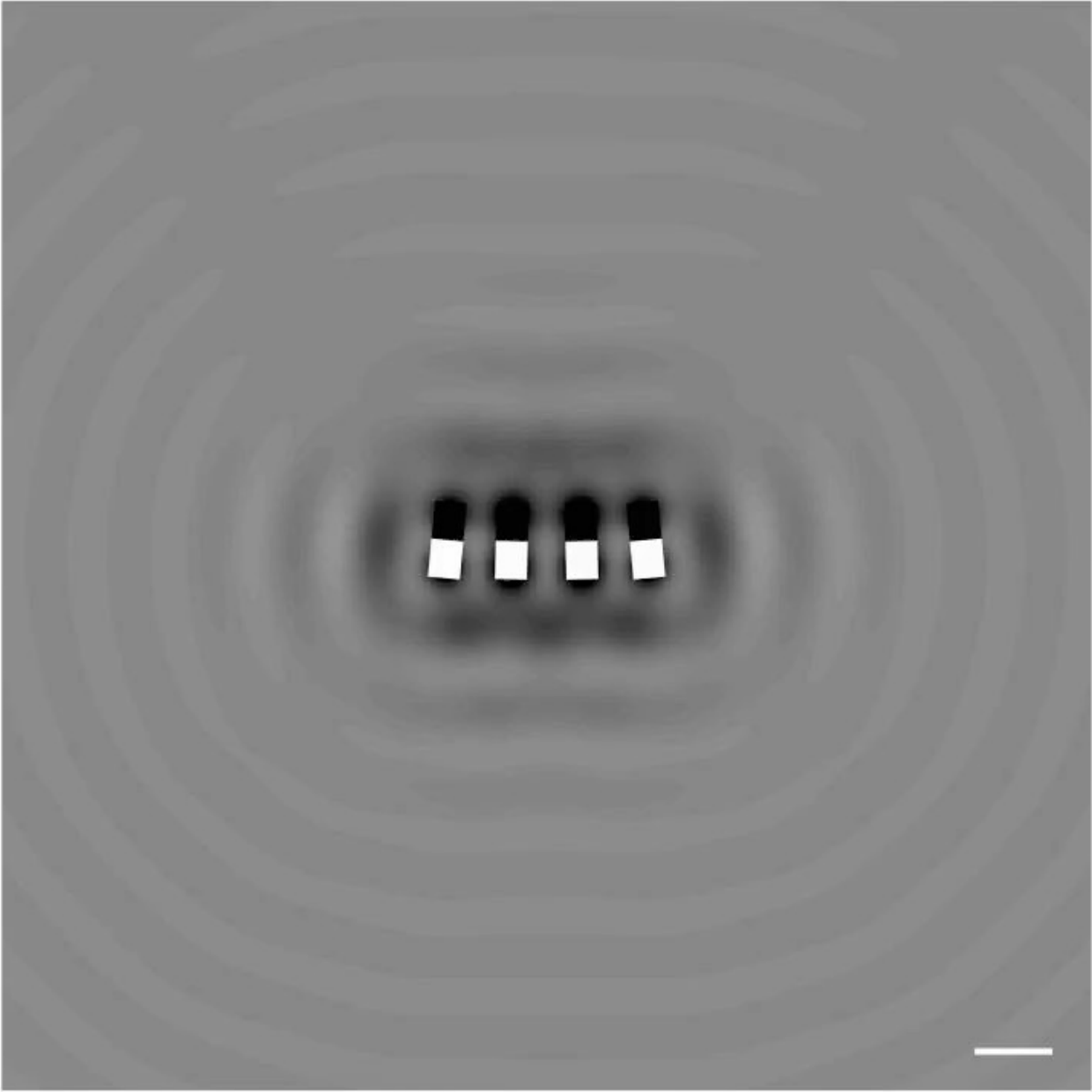} \\ [1mm]
  \raggedright { {\bf Video 6:} A 4-surfer promenade mode and corresponding wavefield for $f = 100$ Hz and $\gamma/g = 3.3$, as shown in Fig.~\ref{CollectivePlot}(a). Neighbors are separated by approximately one capillary wavelength $\lambda_c$, which is indicated by the scale bar.}
\end{figure*}

\begin{figure*}[ht]
  \centering
    \includegraphics[width=0.2\textwidth]{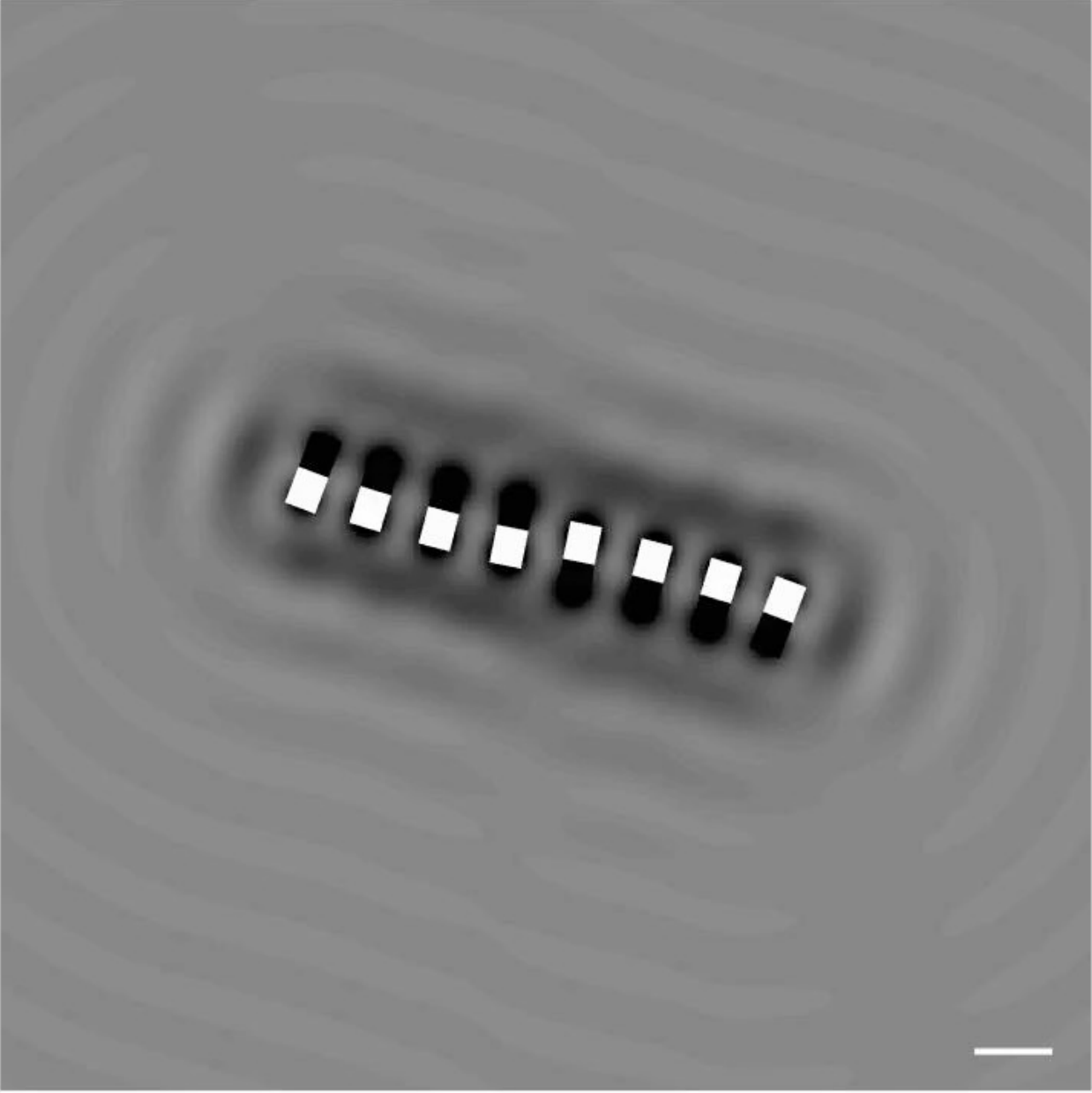} \\ [1mm]
  \raggedright { {\bf Video 7:} An 8-surfer super-orbiting mode and corresponding wavefield for $f = 100$ Hz and $\gamma/g = 3.3$, as shown in Fig.~\ref{CollectivePlot}(b). The scale bars denote the capillary wavelength $\lambda_c$.}
\end{figure*}

\begin{figure*}[ht]
  \centering
    \includegraphics[width=0.2\textwidth]{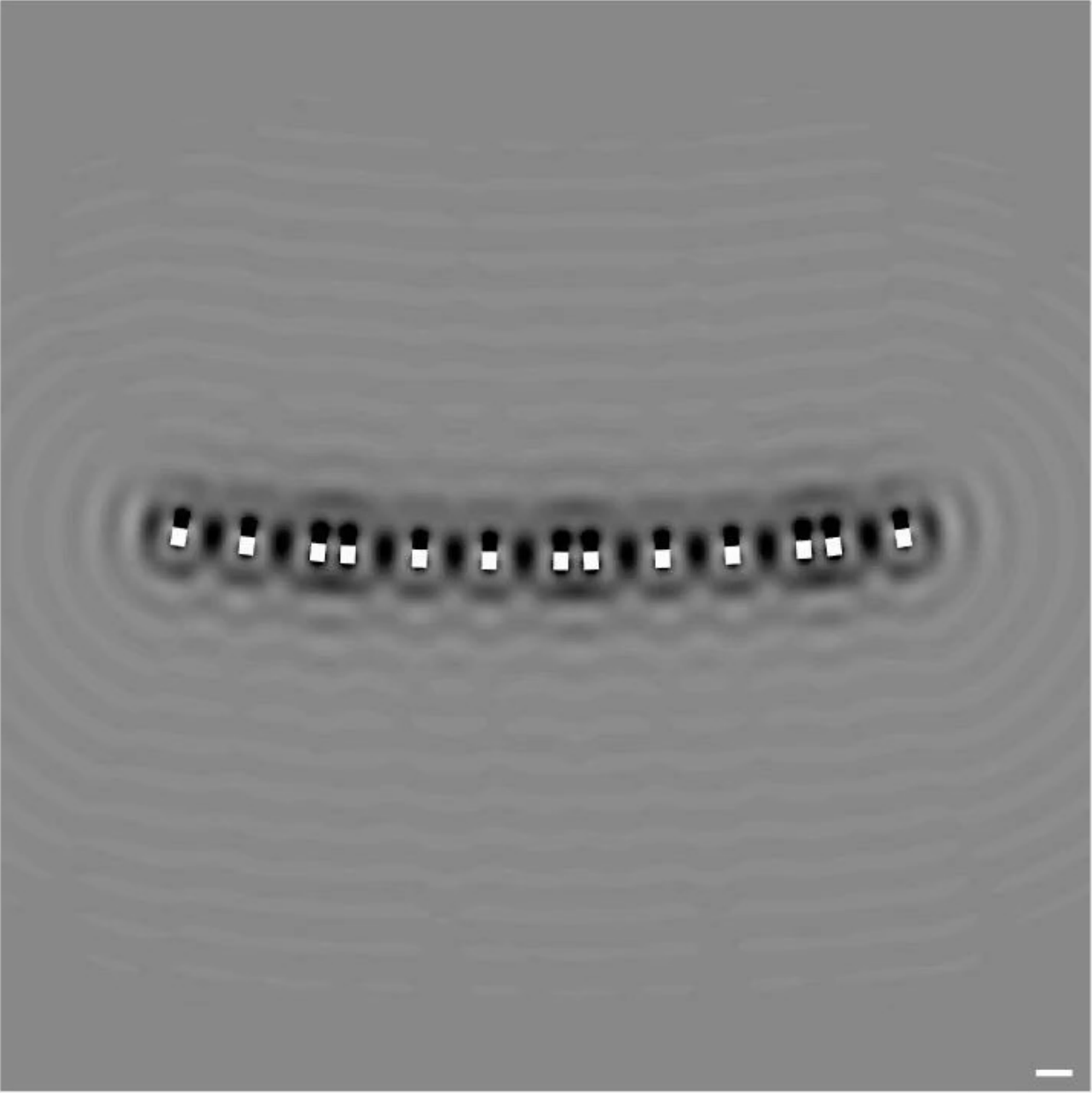} \\ [1mm]
  \raggedright { {\bf Video 8:} A flocking state of thirteen surfers and corresponding wavefield for $f = 100$ Hz and $\gamma/g = 3.3$, as shown in Fig.~\ref{CollectivePlot}(c). Pairs of surfers are separated by approximately one or two capillary wavelengths $\lambda_c$, which is indicated by the scale bar.}
\end{figure*}

\begin{figure*}[ht]
  \centering
    \includegraphics[width=0.2\textwidth]{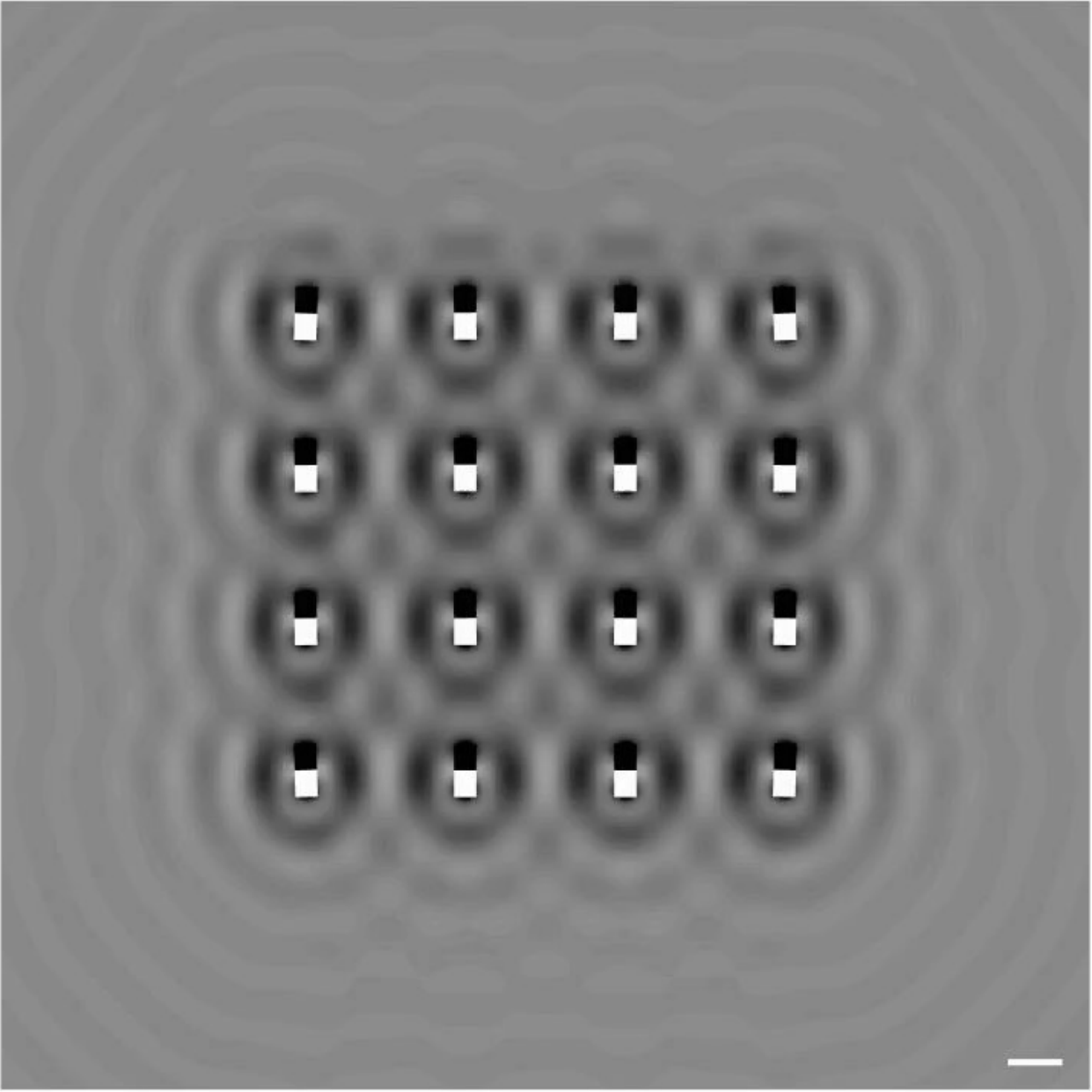} \\ [1mm]
  \raggedright { {\bf Video 9:} A flocking state of sixteen surfers and corresponding wavefield for $f = 100$ Hz and $\gamma/g = 3.3$, as shown in Fig.~\ref{CollectivePlot}(d). Neighboring surfers are separated in both the horizontal and vertical directions by approximately three capillary wavelengths $\lambda_c$, which is indicated by the scale bar.}
\end{figure*}

\bibliography{biblio_surfers,ARFMBib,SchoolingBib}

\end{document}